\documentclass{aa}  
\usepackage{graphicx}
\usepackage{txfonts}
\usepackage{threeparttable}
\usepackage[normalem]{ulem}
\usepackage[version=4]{mhchem} 
\usepackage{float}
\usepackage{placeins}
\usepackage{capt-of}
\usepackage{footnote}
\providecommand{\as}{^{\prime\prime}}
\DeclareUnicodeCharacter{2212}{-}

\usepackage[colorlinks=true,citecolor=blue, urlcolor=blue,
linkcolor=red]{hyperref}
\hypersetup{pageanchor=false}
\usepackage[export]{adjustbox}
\usepackage{etoolbox}
\makeatletter
\patchcmd{\NAT@citex}
 {\@citea\NAT@hyper@{%
    \NAT@nmfmt{\NAT@nm}%
    \hyper@natlinkbreak{\NAT@aysep\NAT@spacechar}{\@citeb\@extra@b@citeb}%
    \NAT@date}}
 {\@citea\NAT@nmfmt{\NAT@nm}%
  \NAT@aysep\NAT@spacechar\NAT@hyper@{\NAT@date}}{}{}

\patchcmd{\NAT@citex}
 {\@citea\NAT@hyper@{%
    \NAT@nmfmt{\NAT@nm}%

\hyper@natlinkbreak{\NAT@spacechar\NAT@@open\if*#1*\else#1\NAT@spacechar\fi}%
      {\@citeb\@extra@b@citeb}%
    \NAT@date}}
 {\@citea\NAT@nmfmt{\NAT@nm}%

\NAT@spacechar\NAT@@open\if*#1*\else#1\NAT@spacechar\fi\NAT@hyper@{\NAT@date}}
 {}{}
\renewcommand*\aa@pageof{, page \thepage{} of \pageref*{LastPage}}
\makeatother

\begin{document} 

   \title{Radial variations in nitrogen, carbon, and hydrogen fractionation in the PDS~70 planet-hosting disk}
   \authorrunning{L. Rampinelli et al.}
   \author{L. Rampinelli \inst{1}, S. Facchini \inst{1}, M. Leemker\inst{1}, P. Curone\inst{2}, M. Benisty \inst{3}, K. I. \"Oberg\inst{4}, R. Teague\inst{5}, S. Andrews\inst{4}, J. Bae \inst{6}, C. J. Law\inst{7,8}, B. Portilla-Revelo\inst{9, 10}}

   \institute{Dipartimento di Fisica, Universit\`a degli Studi di               Milano, Via Celoria 16, 20133 Milano, Italy\\
            \email{luna.rampinelli@unimi.it}
        \and
            Departamento de Astronomía, Universidad de Chile, Camino El Observatorio 1515, Las Condes, Santiago, Chile
        \and
            Max-Planck Institute for Astronomy (MPIA), K\"onigstuhl 17, 69117 Heidelberg, Germany
        \and
            Center for Astrophysics | Harvard \& Smithsonian, 60 Garden St., Cambridge, MA 02138, USA
        \and 
            Department of Earth, Atmospheric, and Planetary Sciences, Massachusetts Institute of Technology, Cambridge, MA 02139, USA
        \and 
            Department of Astronomy, University of Florida, Gainesville, FL 32611, USA
        \and
            NASA Hubble Fellowship Program Sagan Fellow
        \and
            Department of Astronomy, University of Virginia, Charlottesville, VA 22904, USA
        \and 
            Center for Exoplanets and Habitable Worlds, Penn State University, 525 Davey Laboratory, 251 Pollock Road, University Park, PA, 16802, USA
        \and
            Department of Astronomy \& Astrophysics, The Pennsylvania State University, 525 Davey Laboratory, University Park, PA 16802, USA
            }

   \date{Received 18 February 2025; accepted 02 April 2025}
   
 
  \abstract{
Element isotopic ratios are powerful tools to reconstruct the journey of planetary material, from the parental molecular cloud to protoplanetary disks, where planets form and accrete their atmosphere. Radial variations in isotopic ratios in protoplanetary disks reveal local pathways which can critically affect the degree of isotope fractionation of planetary material. In this work we present spatially-resolved profiles of the \ce{^14N}/\ce{^15N}, \ce{^12C}/\ce{^13C}, and D/H isotopic ratios of the HCN molecule in the PDS~70 disk, which hosts two actively-accreting giant planets. ALMA high spatial resolution observations of HCN, \ce{H^13CN}, \ce{HC^15N}, and DCN reveal radial variations of fractionation profiles. We extract the HCN/\ce{HC^15N} ratio out to $\sim$120~au, which shows a decreasing trend outside the inner cavity wall of the PDS~70 disk located at $\sim50$~au. We suggest that the radial variations observed in the HCN/\ce{HC^15N} ratio are linked to isotope selective photodissociation of \ce{N2}. We leverage the spectrally resolved hyperfine component of the HCN line to extract the radial profile of the HCN/\ce{H^13CN} ratio between $\sim$40 and 90~au, obtaining a value consistent with the ISM \ce{^12C}/\ce{^13C} ratio. The deuteration profile is also mostly constant throughout the disk extent, with a DCN/HCN ratio $\sim$0.02, in line with other disk-averaged values and radial profiles in disks around T~Tauri stars. The extracted radial profiles of isotopologue ratios show how different fractionation processes dominate at different spatial scales in the planet-hosting disk of PDS~70.}
    

   \maketitle
%
\section{Introduction} \label{sec:intro}
The process of planet formation is expected to leave imprints on the natal environment, i.e. the protoplanetary disk, such as dust substructures \citep[see e.g.][]{andrews2018disk, sierra2021molecules, benisty2022optical}, kinematic deviations in the rotation pattern of the gas \citep[see e.g.][]{perez2015planet, pinte2022kinematic, izquierdo2023maps}, or chemical signatures \citep[see e.g.][]{cleeves2015indirect, law2023so, booth2023sulphur}. Thanks to the incredible capabilities of both ground based facilities and space missions, a large variety of indirect evidence of ongoing planet formation in protoplanetary disks has been presented, including the first multi-wavelength direct detection of two accreting gas giants in the transition disk around PDS~70 \citep{keppler2018discovery, haffert2019two, benisty2021circumplanetary, zhou2021hubble}. 

The first chemical inventory of the PDS~70 disk \citep{bib:facchini2021chemical} and recent high resolution line emission observations \citep{rampinelli2024} showed a complex structured emission morphology of molecular tracers, highlighting the strong link between forming planets and their natal environment. In particular, most of the molecules show a ring-shaped emission co-located with the cavity wall and the peak in the scattered light from $\mu$m dust \citep{keppler2018discovery}, which is consistent with the presence of a deep gas gap in the PDS~70 disk \citep{keppler2019highly,portilla2023constraining,law2024mapping}, and feeds back into the chemical makeup of the planets that are forming \citep[see e.g.][]{cridland2023planet,leemker2024chemistry,Hsu2024pds70b}. 

\begin{figure*}[ht!]
    \centering
    \includegraphics[scale=0.35]{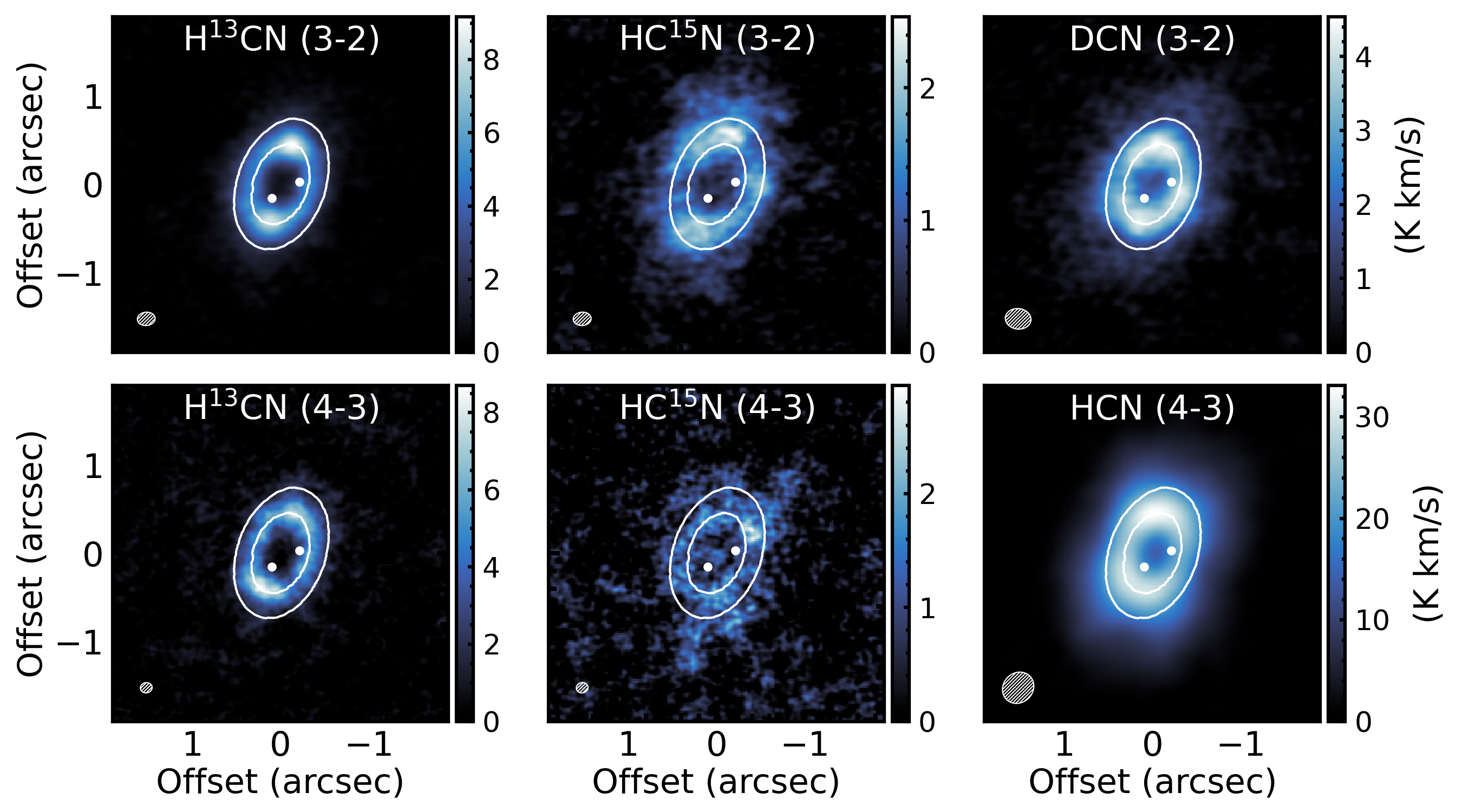}
    \caption{Integrated intensity maps of HCN isotopologues in the PDS~70 disk.  Brightness temperatures were obtained under the assumption of Rayleigh-Jeans approximation. The HCN (4-3) map in the bottom right panel is obtained from SB observations only. The white contours show the bright ring in the sub-mm continuum emission \citep{isella2019detection}, the white dots mark the position of the two forming planets \citep[2020 astrometry data from][]{wang2021constraining}, and the gray ellipse on the bottom left of each panel shows the synthesized beam.}
    \label{fig:M0}
\end{figure*}

Multiple aspects of disk chemistry are linked to the process of planet formation \citep[see e.g.][ and references therein]{oberg2021astrochemistry, oberg2023protoplanetary}, such as location of snowlines \citep{oberg2011effect, okuzumi2019nonsticky}, ratios of the main elements H, C, O, and N \citep{cridland2020connecting, drazkowska2023planet, jiang2023chemical}, or isotopic ratios \citep{altwegg2019cometary, nomura2022isotopic}. The latter can be an efficient tool to trace back the origin of planetary material, as it stores important markings of the history of gas and solids from the molecular cloud, down to rocky cores and planetary atmospheres building in protoplanetary disks \citep{nomura2022isotopic}. Isotopic levels in solids and gas have been used to reconstruct the origin of volatiles in Solar System objects \citep{morbidelli2000source, aleon2010multiple, marty2012earth, ceccarelli2014deuterium, albertsson2014chemodynamical, altwegg2019cometary}. Isotopic ratios of molecules are primarily set when molecules form, but various environmental conditions during the timeline of planet formation can reset or alter these values: this is highlighted by the large range spanned by isotopic ratios in different Solar System bodies, particularly for D/H and \ce{^14N}/\ce{^15N} \citep[see e.g.][and references therein]{nomura2022isotopic}. The first measurement of \ce{^14NH3}/\ce{^15NH3} ratio for a brown dwarf atmosphere was presented by \cite{barrado2023}, showing a value of $\sim670$, which is consistent with star-like formation by gravitational collapse. This result is also carving the path towards direct isotopologue ratios measurements in exoplanetary atmospheres. In particular, a measurement of \ce{^12CO}/\ce{^13CO} has been presented for the first time for exoplanetary atmospheres \citep{zhang202113co, gandhi2023}, showing \ce{^12C/^13C} values lower than the ISM one. 

These recent breakthroughs show the potential of isotopic imprints not only to trace the origins of the solar system, but also of exoplanets. 
In this context, direct constraints on the isotopic ratios in the protoplanetary disk stage, where planets build their cores, and accrete their atmospheres, are fundamental. Disk-integrated isotopologue ratios have been achieved in a variety of molecular tracers, including CO, HCN, CN, \ce{C2H}, and \ce{HCO+} \citep{smith2009high,oberg2012evidence,guzman2015cyanide,zhang2017mass,huang2017alma,guzman2017,hily-blant2017,hily-blant2019multiple, hily2019multiple, yoshida2022, yoshida2024, bergin2024}. These results provided first insights on the main chemical processes driving fractionation in disks: for example, both warm and cold gas-phase deuteration pathways were suggested to take place in disks, while isotope selective photodissociation was proposed to play a crucial role for the nitrogen fractionation (see \citealt{nomura2022isotopic} for a recent review). Moreover, a first comparison with earlier and later stages of star and planet formation revealed the importance of comparing isotopologue ratios extracted from the same molecule, since fractionation processes can lead to very different isotope ratios in different tracers \citep{bergin2024}. Finally, these disk-integrated results already show important deviations from the ISM isotopic ratios.

On the other hand, disk-integrated ratios do not fully capture the chemical complexity of protoplanetary disks, which is expected to induce spatial variations of isotopic ratios as different fractionation processes dominate at different scales and disk regions.
Radially resolved isotopologue ratios were extracted in the MAPS sample for DCN/HCN \citep{cataldi2021molecules}, for \ce{HCN}/\ce{HC^15N} in the TW Hya \citep{hily2019multiple}, and the V4046 Sgr (\citealt{nomura2022isotopic}, Guzmán et al. in prep.) disks, and for \ce{^13CO}/\ce{C^18O} in the TW~Hya disk \citep{furuya2022detection}.

In this work we present the radial profile of nitrogen, carbon, and hydrogen fractionation of the HCN molecule in the PDS~70 disk, where two protoplanets are accreting their atmospheres. In Section~\ref{sec:observations} we present ALMA line emission observations we analyzed. In Section~\ref{sec:column} we describe how we reconstructed the radial profiles of the column density of the HCN isotopologues, and the related isotopologue ratios. We discuss the results in Section~\ref{sec:discussion}, and summarize our conclusions in Section~\ref{sec:conclusion}.

\section{Observations} \label{sec:observations}
\begin{table*}
  \renewcommand{\arraystretch}{1.2}
  \center
  \caption{Listed imaged lines, corresponding rest frequencies, line properties, and imaging parameters (channel width, beam, RMS).}
  \label{tab:imaging_molecules}
  \begin{threeparttable}
    \footnotesize
    \setlength{\tabcolsep}{3pt}
    \tiny
      \begin{tabular}{c c c c c c c c c c c c}
          \hline
          \hline
          Transition\tnote{a} & Frequency & $E_\mathrm{u}$ & log(A$_\mathrm{ul}$) & $g_\mathrm{u}$ & Q(18.75K) & Q(37.50K) & Q(75K) & $\Delta \mathrm{v}$ & Beam & RMS\tnote{b} & $\Delta \mathrm{v}_{\mathrm{M}0}$\tnote{c} \\
          & [GHz] & [K] & [s$^{-1}$] & & & & & [km~s$^{-1}$] & bmaj $\times$ bmin (PA) &[mJy~beam$^{-1}$] & [km~s$^{-1}$] \\
          \hline
          H$^{13}$CN (3-2) & 259.0117 & 24.9 & $-3.11$ & 21 & 28.17 & 55.31 & 109.6 & 0.4 & $0\farcs20 \times 0\farcs15 \,(95^\circ)$& 0.55 & 1.5 - 9.5\\
          H$^{13}$CN (4-3) & 345.3398 & 41.4 & $-2.72$ & 27 & 28.17 & 55.31 & 109.6 & 0.4 & $0\farcs13 \times 0\farcs12\,(100^\circ)$ & 0.70 & 1.85 - 9.5\\
          HC$^{15}$N (3-2) & 258.1571 & 24.8 & $-3.11$ & 7 & 9.421 & 18.50 & 36.66 & 0.4 & $0\farcs20 \times 0\farcs15\,(94^\circ)$ & 0.65 & 2.3 - 8.7\\
          HC$^{15}$N (4-3) & 344.2001 & 41.3 & $-2.73$ & 9 & 9.421 & 18.50 & 36.66 & 0.9 & $0\farcs13 \times 0\farcs12\,(101^\circ)$ & 0.75 & 1.9 - 9.1\\
          DCN (3-2) & 217.2386 & 20.9 & $-3.33$ & 21 & 33.39 & 65.75 & 130.5 & 0.9 & $0\farcs29 \times 0\farcs23\,(83^\circ)$ & 0.73 & 2.3 - 9.1\\
          HCN ($J$=4-3,$F$=4-4) & 354.50387 & 42.5 & $-3.89$ & 9 & 27.47 & 53.91 & 106.8 &  &  &  & \\   
          HCN ($J$=4-3,$F$=3-2) & 354.50537 & 42.5 & $-2.72$ & 7 & 27.47 & 53.91 & 106.8 &  &  &  & \\
          HCN ($J$=4-3,$F$=4-3)\tnote{d} & 354.50548 & 42.5 & $-2.72$ & 9 & 27.47 & 53.91 & 106.8 & 0.9 & $0\farcs37 \times 0\farcs34\,(134^\circ)$ & 1.03 & 0.1 - 10 \\
          HCN ($J$=4-3,$F$=5-4) & 354.50552 & 42.5 & $-2.69$ & 11 & 27.47 & 53.91 & 106.8 &  &  &  & \\
          HCN ($J$=4-3,$F$=3-4) & 354.50585 & 42.5 & $-5.58$ & 7 & 27.47 & 53.91 & 106.8 &  &  &  & \\
          HCN ($J$=4-3,$F$=3-3)\tnote{e} & 354.50745 & 42.5 & $-3.78$ & 7 & 27.47 & 53.91 & 106.8 &  &  & \\
          \hline
          \hline
      \end{tabular}
      \begin{tablenotes}
          \item[a] Quantum numbers are formatted as $J$ \citep[CDMS, ][]{fuchs2004high, cazzoli2005lamb, maiwald2000pure,fuchs2004high, cazzoli2005lamb,brunken2004sub,ahrens2002HCN}.
          \item[b] Cubes are imaged with natural weighting.
          \item[c] Velocity ranges over which data cubes where collapsed to extract integrated intensity, chosen after visual inspection of the data. 
          \item[d] Channel width, beam size, and RMS are indicated only for the main HCN component for reference. 
          \item[e] Only HCN hyperfine component spectrally resolved by the presented observations.  
      \end{tablenotes}
      \end{threeparttable}
\end{table*} 
In this work we used ALMA Band 6 and 7 line emission observations from projects \#2019.1.01619.S (PI S. Facchini) and \#2022.1.01695.S (PI M. Benisty). The observations include various lines of HCN isotopologues: HCN ($4-3$), \ce{H^13CN} ($3-2$), \ce{H^13CN} ($4-3$), \ce{HC^15N} ($3-2$), \ce{HC^15N} ($4-3$), DCN ($3-2$).

The Band 7 observations consist of four execution blocks (EBs): one short-baseline (SB, ALMA configuration C-4, max baseline 0.8 km) and three long-baseline (LB, configuration C-7, max baseline 3.6 km) configurations. The spectral setup includes four spectral windows (spws) initially intended for continuum observations, each with a bandwidth of 1.875 GHz, 1920 channels, and a resolution of 977 kHz. The spws cover the same frequency ranges across all EBs: 342.483-344.358~GHz for spw~0, 344.379-346.254~GHz for spw~1, 354.483-356.358~GHz for spw~2, and 356.441-358.316~GHz for spw~3. Each EB had 35 minutes of on-source time, for a total of 2.33 hours. All the EBs were initially calibrated using the ALMA pipeline, followed by self-calibration based on the pipeline designed by the exoALMA Large Program (Loomis et al., subm.), using the software CASA \citep{CASA2022}, version 6.2. We first flagged the lines present in our frequency range (see Table~\ref{tab:lines_b6b7}): \ce{CS} ($7-6$, $\nu=342.88\,\mathrm{GHz}$), \ce{HC^15N} ($4-3$, $\nu=344.20\,\mathrm{GHz}$), \ce{H^13CN} ($4-3$, $\nu=345.34\,\mathrm{GHz}$), \ce{^12CO} ($3-2$, $\nu=345.80\,\mathrm{GHz}$), \ce{HCN} ($4-3$, $\nu=354.51\,\mathrm{GHz}$), \ce{HCO^+} ($4-3$, $\nu=356.73\,\mathrm{GHz}$), and \ce{SO} ($8_8-7_7$, $\nu=344.31\,\mathrm{GHz}$). We flagged the range between -30 km/s to +30 km/s around each line center. The remaining unflagged data were averaged into 250~MHz-wide channels. A first round of phase-only self-calibration was performed on each EB separately, combining both scans and spws on time intervals over the full EB duration. The EBs were then aligned by regridding them to a common uv-grid with natural weighting, retaining only overlapping grid cells, and applying phase shifts to minimize visibility differences (Loomis et al., subm.). No flux rescaling was necessary, as flux offsets between EBs were minimal ($<1\%$). Next, phase-only self-calibration was applied to the SB~EB using models created with the task \texttt{tclean}, cleaning down to $6\sigma$ at each round. Multiple rounds with progressively shorter solution intervals were used (EB-long, 360s, 120s, 60s, 20s), stopping when SNR and noise structure began to degrade. The self-calibrated SB~EB was then concatenated with the three LB~EBs, and phase-only self-calibration was repeated on the combined dataset. Similar to before, we cleaned down to $6\sigma$ at each step, with progressively shorter solution intervals (EB-long, 360s, 120s, 60s, 30s, 18s). After this, the data were cleaned down to $1\sigma$, and two rounds of amplitude and phase self-calibration were performed, combining polarizations and spws. The first round used EB-long intervals, while the second used scan-long intervals. Gain solutions and phase shifts were then applied to the full spectral data including line emission channels, following exactly the same order. To reduce dataset size, the full spectral data were binned into 30s intervals. Finally, continuum subtraction was performed on all datasets using the \texttt{uvcontsub} task. Self-calibration of Band~6 data was performed as described in  \cite{law2024mapping,rampinelli2024}.

We imaged the lines using the \texttt{tclean} task \citep{hogbom1974aperture,cornwell2008} implemented in the \texttt{CASA} software \citep{CASA2022} \texttt{v6.65}. In particular, we used a \texttt{multiscale} deconvolver with scales [0,5,10,20,30] pixels, where the pixel size is $\sim0\farcs02$, and a flux threshold of 3.5$\sigma$. We performed a first CLEANing without applying any mask, and a second CLEANing iteration using a mask built by selecting the signal $> 7\sigma$ in the previous iteration (Loomis et al., subm.). We applied the so-called JvM correction  \citep{jorsater1995high}, following \cite{czekala2021molecules}, by rescaling the residual map by the ratio between areas of the CLEAN and DIRTY beams ($\epsilon$), to correct for the units mismatch between the model and residuals image (see Table~\ref{tab:imaging_molecules_2} for the $\epsilon$ values corresponding to each cube). More details on the CLEANing strategies can be found in \cite{rampinelli2024}. The following analysis was performed on continuum subtracted cubes, and JvM corrected cubes, except for uncertainty evaluation, which could be underestimated by the JvM correction \citep{casassus2022variable}, as residuals are rescaled. In the following analysis we also assumed the geometrical parameters\sout{,} $M_\ast = 0.875$ M$_\odot$, $i = 51.7^\circ$, PA$=160.4^\circ$, and $v_\mathrm{sys}=5.5$~km~s$^{-1}$ (in LSRK frame), as they were obtained by \cite{keppler2019highly}, and a distance of $113$ pc \citep{brown2018gaia}.
 
We first imaged all lines with natural weighting. Table~~\ref{tab:imaging_molecules} lists the image properties, including channel width, beam size, and RMS, and properties of the molecular transitions. RMS is evaluated as standard deviation in the first and last 5 channels of each cube. A list of detected lines, with corresponding line fluxes is listed in Table~\ref{tab:lines_b6b7}, which also includes Band~6 lines previously presented by \cite{rampinelli2024}, for comparison. Line fluxes where obtained as described in Appendix~\ref{appendix:fluxes} (Band~7) and \cite{rampinelli2024} (Band~6).

In order to perform the analysis outlined in Sect.~\ref{sec:column}, we imaged all the cubes with two different sets of parameters to maximize the sensitivity of Band~6 and 7 observations, respectively. Each time $uv$-taper and/or robust were chosen to obtain cubes with the same spatial resolution. In particular, cubes imaged to maximize Band~6 and Band~7 sensitivity correspond to beam major axes of $\sim0\farcs20$ ($\sim$23 au) and $\sim0\farcs14$ ($\sim$15 au), respectively. Corresponding weighting, $uv$-taper, beam size, beam area, and RMS for the two imaging rounds are listed in Table~\ref{tab:imaging_molecules_2}.

We calculated integrated intensity maps of the imaged cubes using the \texttt{bettermoments} package \citep{teague2019bettermoments}, integrating data cubes over a velocity range where each line shows emission in the channel maps. Figure~\ref{fig:M0} shows an example of the obtained maps for each line (see last column of Table~\ref{tab:imaging_molecules}). The map in the bottom right panel of Fig.~\ref{fig:M0} is obtained from SB observations only, which completely cover the HCN (4-3) line, while LB observations spectrally cover only half of the line, allowing the reconstruction of half of the azimuthal extent of the disk. In this work we analyzed the cube covering half of the HCN (4-3) line, obtained from combined SB and LB observations, as high angular resolution is needed to separate the hyperfine component (see Appendix~\ref{appendix:imaging} for more details). 

\section{Results} \label{sec:column}
\subsection{Radial rotational diagram}
\begin{figure*}[ht!]
    \centering
    \includegraphics[scale=0.17]{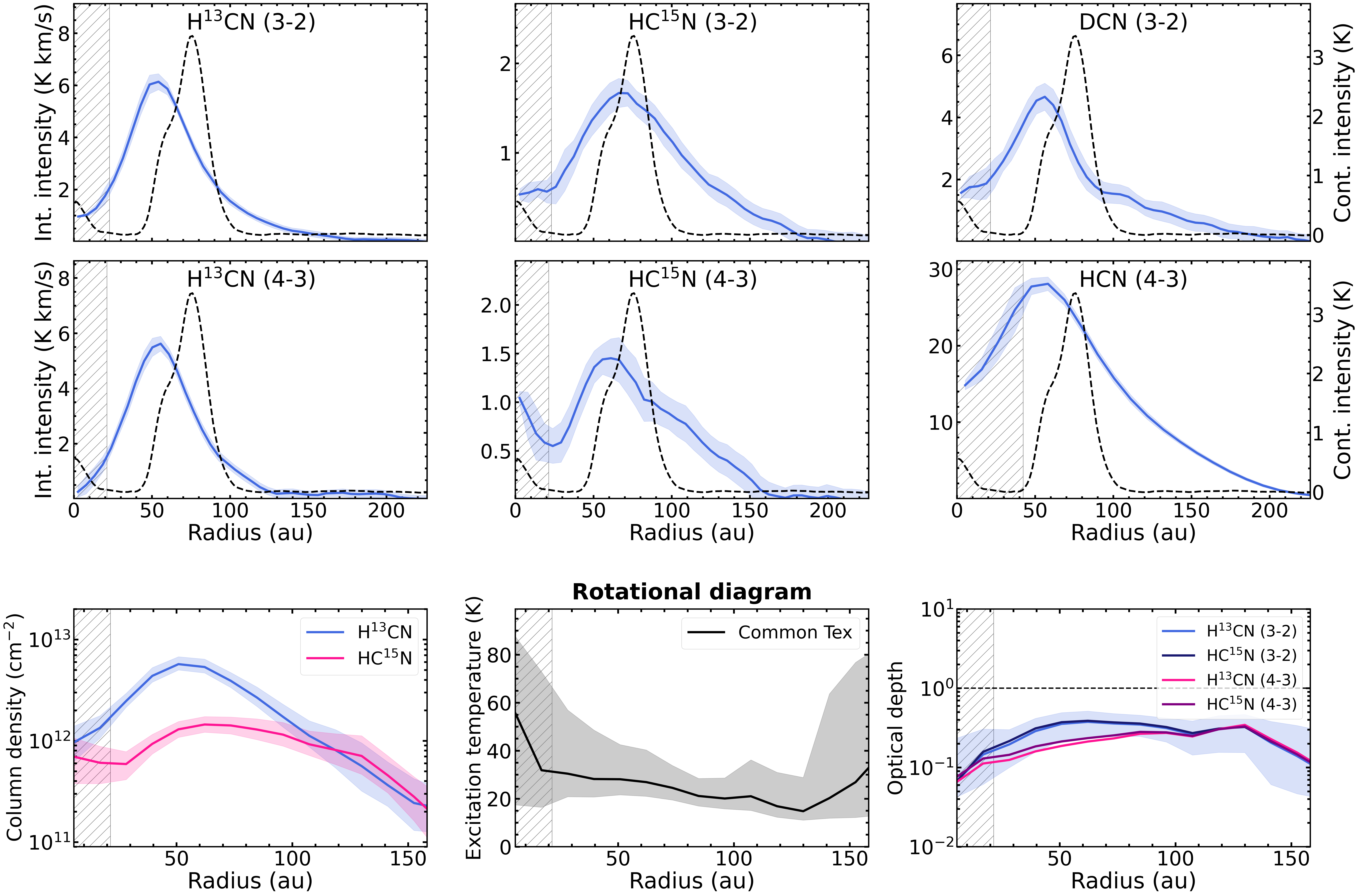}
    \caption{Top panels: integrated intensity profiles of lines of HCN isotopologues. The ribbons show the standard deviation across each annulus divided by the square root of the number of independent beams. The intensity is expressed in brightness temperature under the Rayleigh-Jeans approximation. The black dashed line is the integrated intensity profile of the 855~$\mu$m continuum \citep{isella2019detection, benisty2021circumplanetary}. The hatched region shows the beam major axis of each line. Bottom row: results of the radial rotational diagram analysis. Solid lines in the left and middle panels show the 50th percentile of column density profiles (\ce{H^13CN} in blue and \ce{HC^15N} in pink), and of the excitation temperature (black). Ribbons show the 16th and 84th percentiles of the posterior distributions. The right panel represents the radial profiles of the optical depth for the analyzed molecular lines of \ce{H^13CN} and \ce{HC^15N}, and the related uncertainty shown for the \ce{H^13CN} (3-2) line, for reference, as it is representative of the typical uncertainty. The dashed black line shows the $\tau=1$ level.}
    \label{fig:profiles_rotdiagr}
\end{figure*}

From the integrated intensity maps in Fig.~\ref{fig:M0}, we extracted the radial profiles of the integrated intensity of each line, using the \texttt{GoFish} package \citep{teague2019gofish}. In order to limit beam dilution when performing the azimuthal average, we tried to extract the radial profiles of the integrated intensity only from a wedge of $\pm30^\circ$ along the disk major axis. However, the resulting signal-to-noise ratio for DCN and \ce{HC^15N} in particular was too low to allow for the analysis presented in this paper. We therefore extracted the integrated intensity profiles by performing a complete azimuthal average over $0\farcs1$ wide annuli. The resulting profiles are plotted in the top panels in Fig.~\ref{fig:profiles_rotdiagr}. The dashed black line shows the integrated intensity profile of the 855~$\mu$m continuum observations \citep[][corresponding beam major axis of $0\farcs05$]{isella2019detection, benisty2021circumplanetary}. 

As two lines are detected for both \ce{HC^15N} and \ce{H^13CN}, with upper energies ranging from $\sim25$ to 41~K, we could apply the rotational diagram \citep{goldsmith1999population} analysis to extract their molecular column density $N_\mathrm{t}$ and excitation temperature $T_\mathrm{ex}$, as follows: 
\begin{equation}\label{eq:rot_diag}
    \ln{\frac{N_\mathrm{u}}{g_\mathrm{u}}} = \ln{N_\mathrm{t}} -\ln{Q(T_\mathrm{ex})} - \frac{E_\mathrm{u}}{\mathrm{k_\mathrm{B}}T_\mathrm{ex}},
\end{equation}
where $N_\mathrm{u}$, $g_\mathrm{u}$, and $E_\mathrm{u}$ are the upper state column density, degeneracy, and energy, respectively, $N_\mathrm{t}$ is the total molecular column density, k$_\mathrm{B}$ is the Boltzman constant, $T_\mathrm{ex}$ is the excitation temperature, and $Q(T_\mathrm{ex})$ is the partition function. In the optically thin assumption, $N_\mathrm{u}$ depends on the integrated flux density $S_\nu \Delta \mathrm{v}$:
\begin{equation} \label{eq:Nu}
    N_\mathrm{u} = \frac{4 \pi S_\nu \Delta \mathrm{v}}{A_\mathrm{ul} \Omega \mathrm{hc}},
\end{equation}
where $\Omega$ is the emitting area.

If the optically thin assumption is not valid, i.e. $\tau \not\ll 1$, Eq.~\ref{eq:rot_diag} can be corrected for the optical depth $\tau$ through the following factor:
\begin{equation}\label{eq:Ctau}
    C_\tau = \frac{\tau}{1-e^{-\tau}},
\end{equation}
resulting in the following rotational diagram equation \citep[see e.g.][]{loomis2018distribution}:
\begin{equation} \label{eq:Ctau_rotdiag}
        \ln{\frac{N_\mathrm{u}}{g_\mathrm{u}}} = \ln{N_\mathrm{t}} - \ln{C_\tau} -\ln{Q(T_\mathrm{ex})} - \frac{E_\mathrm{u}}{\mathrm{k_\mathrm{B}}T_\mathrm{ex}}.
\end{equation}

We computed the rotational diagram correcting for the optical depth factor $C_\tau$. We applied a radial rotational diagram, to retrieve the radial profiles of $N_\mathrm{t}$ and $T_\mathrm{ex}$ for the \ce{H^13CN} and \ce{HC^15N} molecules, using Eq.~\ref{eq:Nu} and~\ref{eq:Ctau_rotdiag}, and cubes with the same beam sizes.

We accounted for the optical depth of each line included in the rotational diagram, evaluating $\tau$ as follow:
\begin{equation}\label{eq:tau}
    \tau(R) = - \ln{\biggl(1-\frac{T_\mathrm{b}(R)}{T_\mathrm{ex}(R)}\biggr)},
\end{equation}
where $T_\mathrm{b}(R)$ is the radial profile of the brightness temperature, taken as the azimuthal average of the peak brightness temperature over each annulus in which we applied the rotational diagram, and $T_\mathrm{ex}(R)$ is the radial profile of the sampled excitation temperature. 
Another important limitation of using the peak brightness temperature comes from the finite spectral resolution of the observations, and thus in spectral smearing which results in underestimating the peak brightness temperature in a specific velocity bin. We therefore used cubes imaged with the smallest width of the velocity channel allowed by the spectral resolution of the observations. We explored the effect of spectral smearing on the peak brightness temperature in Appendix~\ref{appendix:B}, which we evaluated to be negligible in our case, and thus did not include in the analysis.

We sampled the posterior distribution of $T_\mathrm{ex}(R)$ and $N_\mathrm{t}(R)$ through a Markov-Chain Monte Carlo (MCMC) method with the \texttt{emcee} package \citep{foreman2013emcee}, with 128 walkers, 3000 burn-in steps, and 500 steps. We considered a uniform prior with $T_\mathrm{ex}\in (10,100)$~K and $N_\mathrm{t} \in (10^{10}, 10^{16})$~cm$^{-2}$. 
The propagated uncertainty on the integrated intensity, on the brightness temperature and a 10\% absolute flux calibration uncertainty were considered (sum in quadrature) when generating the likelihood. Moreover, we sampled the posterior distribution of the optical depth starting from random samples from the $T_\mathrm{ex}$ posterior distribution, and from a normal distribution of $T_\mathrm{b}$, centered around its median value, with standard deviation given by its statistical uncertainty. 

We first applied the rotational diagram separately to \ce{H^13CN} and \ce{HC^15N}, since two lines are available for each of the molecules. 
As the excitation temperatures obtained for the two molecules are consistent, we applied a final rotational diagram analysis fitting for a single excitation temperature for both \ce{HC^15N} and \ce{H^13CN}. The bottom panels in Fig.~\ref{fig:profiles_rotdiagr} show the results of the radially-resolved rotational diagram, including $N_\mathrm{t}$, $T_\mathrm{ex}$, and $\tau$. The solid lines show the 50th percentile of the resulting distributions, while ribbons show the 16th and 84th percentiles.

Figure~\ref{fig:profiles_rotdiagr} shows the result obtained from cubes imaged maximizing the sensitivity of Band~6 observations, while Appendix~\ref{appendix:C} shows the same result but from cubes imaged to maximize the sensitivity of Band~7 observations, with the latter corresponding to a higher spatial resolution ($\sim0\farcs15$) with respect to the former ($\sim0\farcs20$). Imaging procedures and details on the resulting imaged cubes are presented in Appendix~\ref{appendix:imaging} and Table~\ref{tab:imaging_molecules_2}. The two results are consistent, and were both used to extract the radial profiles of N, C, and H fractionation of HCN, as described in the following Section. 

\subsection{Isotopologue ratios} from fixed \texorpdfstring{$^{12}C/^{13}C$}{12C/13C}\label{subsec:fixed12C13C}

\begin{figure}[ht!]
    \includegraphics[scale=0.35]{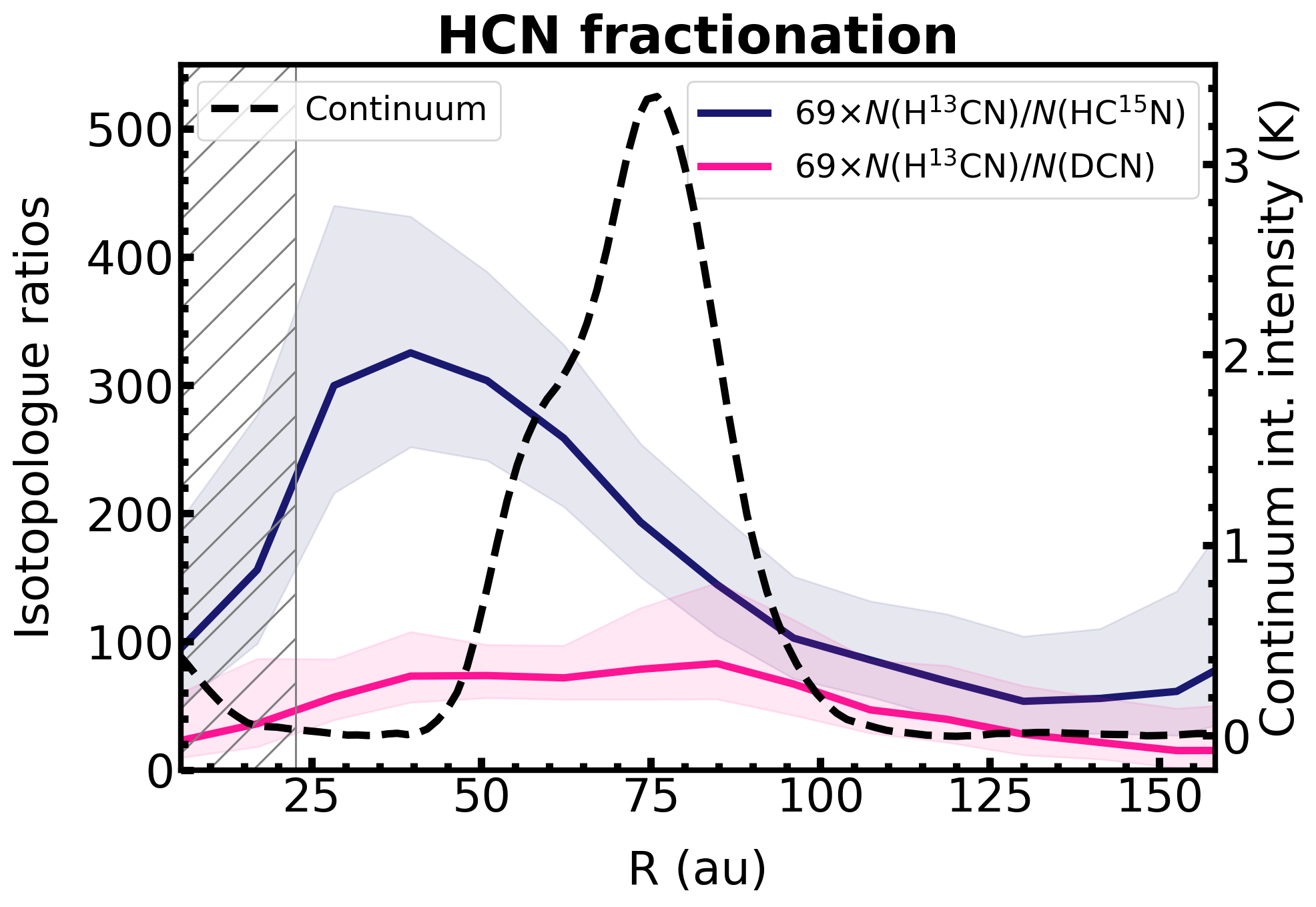}
    \caption{Nitrogen (blue) and hydrogen (pink) fractionation profiles of HCN in the PDS~70 disk, assuming a radially constant $\ce{^12C}/\ce{^13C}=69$ value to convert the \ce{H^13CN} column density into an HCN column density. The ribbons indicate the 16th and 84th percentiles of the posterior distribution of the ratios. The dashed black line shows the integrated intensity profile of the 855~$\mu$m continuum observations \citep{isella2019detection, benisty2021circumplanetary}. The hatched regions indicate the beam major axis of the datacubes of HCN isotopologues.}
    \label{fig:N_D_frac}
\end{figure}

DCN/HCN and HCN/\ce{HC^15N} ratios in disks are usually extracted from \ce{H^13CN} instead of the main isotopologue HCN, as the latter is typically optically thick \citep[see e.g.][]{guzman2017}.
We also extracted the deuteration and nitrogen fractionation profiles by assuming a radially constant $\ce{^12C}/\ce{^13C}=69$ ratio \citep{wilson1999isotopes} to convert the \ce{H^13CN} into an HCN column density. We therefore used the results of the rotational diagram analysis shown in Fig.~\ref{fig:profiles_rotdiagr}, obtained from cubes maximizing the sensitivity of Band~6 observations (see the second section of Table~\ref{appendix:imaging} for imaging specifics). 

The posterior distributions of the $69\times N(\ce{H^13CN})/N(\ce{HC^15N})$ and $N(\ce{DCN})/[69\times N(\ce{H^13CN})]$ ratios were built from the posterior distributions of the column density of \ce{H^13CN}, \ce{HC^15N}, and DCN. We reconstructed the posterior distribution of $N$(DCN) from Eq.~\ref{eq:Ctau_rotdiag}, by randomly sampling the excitation temperature from the rotational diagram result, assuming DCN has the same excitation temperature as \ce{H^13CN} and \ce{HC^15N}. The uncertainty on the DCN flux was accounted for in the posterior distribution of the isotopologue ratio, by randomly sampling the flux of the DCN (3-2) line from a normal distribution centered around the median value, with a standard deviation given by its statistical uncertainty. As for \ce{H^13CN} and \ce{HC^15N}, we extracted the radial profile of the DCN column density correcting for its optical depth, using Eq.~\ref{eq:tau}. The bottom row in Fig.~\ref{fig:profiles_rotdiagr_b7res} in Appendix~\ref{appendix:C} shows the radial profiles of the DCN column density and optical depth, compared to the other isotopologues.

The resulting radial profiles of the isotopologue ratios are presented in Fig.~\ref{fig:N_D_frac}, and are also compared to the radial profile of the continuum integrated intensity (dashed dark line), which do not show any clear correlation with the fractionation profiles.
The nitrogen fractionation profile is decreasing and below the ISM value from $\sim40$ to 100~au, while the deuteration profile is mostly constant between $\sim40$ au 100~au, and significantly above the ISM D/H value of $\sim10^{-5}$ \citep{linsky2006}.

\subsection{Radial profile of HCN column density}\label{subsec:HCN_column}

\begin{figure*}[ht!]
    \includegraphics[scale=0.21]{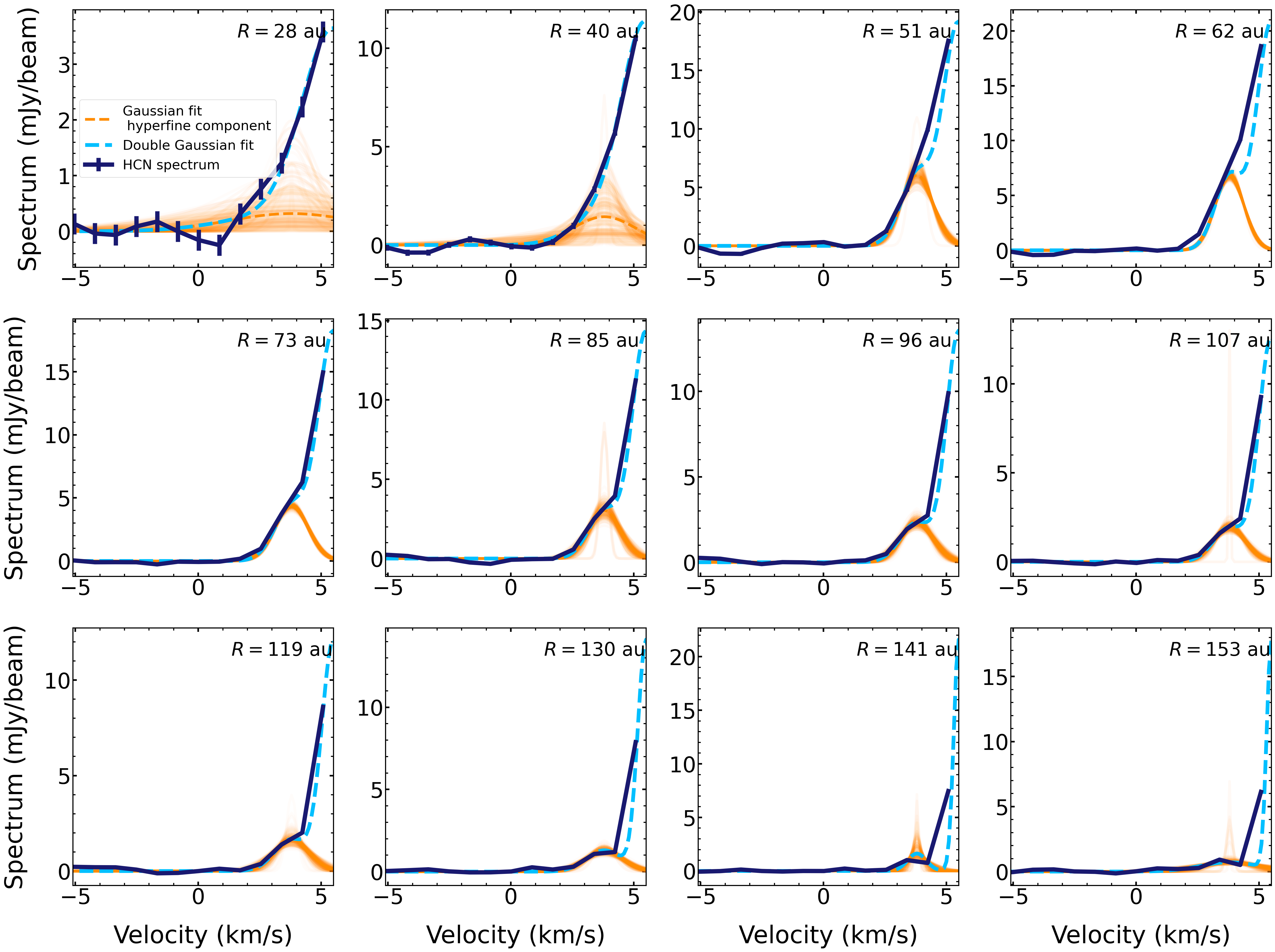}
    \caption{Spectra of the HCN (4-3) line (dark blue lines) at different radii (indicated on the top right of each panel), with the related uncertainties. The dashed light blue line shows the fit to the spectrum, as the sum of two Gaussian components. The orange profiles show the Gaussian fit to the hyperfine component, with the dashed orange line corresponding to the 50th percentile of the posterior distribution, and the solid orange lines represent 100 random samples of the posterior distribution.}
    \label{fig:hyperfine_fit}
\end{figure*}

The radial profile of the HCN column density is needed to extract the HCN/\ce{H^13CN} ratio, or the HCN/\ce{HC^15N} and DCN/HCN ratios without assuming a fixed \ce{^12C/^13C} ratio. HCN column density, optical depth, and excitation temperature could be retrieved leveraging the HCN hyperfine components, which are needed as the main component is typically optically thick, as previously presented by \cite{guzman2021molecules} for the MAPS sample, or by \cite{hily2019multiple} for the TW~Hya disk. 

However, the spatial resolution of SB observations of the HCN (4-3) line does not allow to separate the two hyperfine components, since beam smearing leads to spatial blending of the hyperfine components. LB+SB observations of the analyzed data, on the other hand, allow to spatially resolve the hyperfine component at 354.5075~GHz, centered at 3.8~km~s$^{-1}$ (main component $v_\mathrm{sys}=5.5$~km~s$^{-1}$). As LB observations spectrally cover only one of the two hyperfine components, and only half of the main component (see Fig.~\ref{fig:hyperfine_fit}), we could not apply the rotational diagram to simultaneously extract HCN column density and excitation temperature. We therefore computed the radial profile of the HCN column density, assuming the same excitation temperature obtained from the rotational diagram of \ce{H^13CN} and \ce{HC^15N}, through Eq.~\ref{eq:rot_diag}. 

In particular, we reconstructed the HCN column density from the flux of the hyperfine component at 3.8~km~s$^{-1}$ by assuming it is optically thin, which is valid since $\tau\lesssim0.02$ from Eq.~\ref{eq:tau}. To extract the radial profile of the flux of the hyperfine component, we applied a double-Gaussian fit to the spectrum in multiple radial bins. We fixed the center of the two Gaussian components to be at 3.8~km~s$^{-1}$ (hyperfine component) and at the systemic velocity of 5.5~km~s$^{-1}$ (main component), and we fit for the two widths and peaks. From the result of the Gaussian fit presented in Fig.~\ref{fig:hyperfine_fit}, we then reconstructed the posterior distribution of the hyperfine flux, as outlined in Appendix~\ref{appendix:C} in more details. 

We then reconstructed the posterior distributions of $N$(HCN) using Eq.~\ref{eq:rot_diag} and \ref{eq:Nu}, from the reconstructed posterior distributions of the flux of the HCN hyperfine component, and of the excitation temperature of \ce{H^13CN} and \ce{HC^15N}. In particular, we randomly sampled the excitation temperature from the rotational diagram result, and the hyperfine flux from the posterior distribution obtained from the double-Gaussian fit. 

We were unable to smooth the HCN cube to match the spatial resolution of $\sim0\farcs20$ of \ce{H^13CN} and \ce{HC^15N} cubes used in the rotational diagram shown in Fig.~\ref{fig:profiles_rotdiagr}, as a worse spatial resolution would deteriorate the hyperfine fitting procedure, due to spectral blending of the hyperfine components, as we previously outlined. We therefore used the result from the rotational diagram shown in Appendix~\ref{appendix:C}, obtained from cubes with a higher spatial resolution of $\sim0\farcs15$ (see also Appendix~\ref{appendix:imaging}). 
The bottom left panel of Fig.~\ref{fig:profiles_rotdiagr_b7res} in Appendix~\ref{appendix:C} shows the radial profiles of the HCN column density, in comparison to the other isotopologues.

\subsection{Isotopologue ratios using HCN column density} \label{subsec:vary12C/13C}

From the HCN column density retrieved from the hyperfine fitting procedure, we extracted the radial profiles of the isotopologue ratios HCN/\ce{HC^15N}, HCN/\ce{H^13CN}, and DCN/HCN, shown in Fig.~\ref{fig:12C13C}, by the light blue, red, and purple profiles, respectively.
Similarly to the previous case where we fixed the \ce{^12C/^13C} ratio, the posterior distributions of the ratios were built from the posterior distributions of the column density of \ce{H^13CN}, \ce{HC^15N} (taken from the rotational diagram result shown in Fig.~\ref{fig:profiles_rotdiagr_b7res}), DCN, and HCN. As we already outlined in Sect.~\ref{subsec:HCN_column},
we used cubes imaged to maximize the sensitivity of Band~7 observations (see the first section of Table~\ref{tab:imaging_molecules_2} in Appendix~\ref{appendix:imaging} for imaging specifics), as a higher angular resolution is needed to perform the hyperfine fitting procedure.

As shown in the middle panel of Fig.~\ref{fig:12C13C}, the carbon fractionation profile is consistent with the ISM value between $\sim50$ and 100~au, within the errorbars, which also justifies the previous assumption we made of a radially constant \ce{^12C/^13C} to extract the deuteration and nitrogen fractionation profiles in Sect.~\ref{subsec:fixed12C13C}, from the \ce{H^13CN} isotopologue. Moreover, the isotopologue ratios obtained by assuming a radially constant $\ce{^12C/^13C}=69$ (dark blue and pink profiles in Fig.~\ref{fig:12C13C}), or from the radial profile of the HCN column density directly (light blue and purple profiles), are consistent within the errorbars.

The hyperfine fitting procedure presented above and in Appendix~\ref{appendix:C} is strongly limited by the fact that only half of the HCN (4-3) line is covered by LB observations, by the low spectral resolution of $\sim0.9$~km~s$^{-1}$, and the lower signal to noise ratio. This consequently results in large error bars for the profiles of isotopologue ratios obtained from the main isotopologue HCN, and does not allow to robustly extract the ratios inside $\sim40$~au and beyond $\sim100$~au.

\section{Discussion} \label{sec:discussion}
\subsection{Nitrogen} \label{subsec:disc:N}
Isotope fractionation in disks is mainly driven by two processes: isotope exchange reactions and isotope selective photodissociation \citep[][and references therein]{nomura2022isotopic}. For the specific case of nitrogen fractionation, the former can enhance the abundance of \ce{N^15N} and \ce{HC^15N} at low temperatures $T\lesssim20$~K \citep{terzieva2000possibility}.
On the other hand, isotope selective photodissociation of \ce{N^15N} over the self-shielding \ce{N2} is favored in strongly irradiated regions \citep{heays2014isotope, visser2018nitrogen}, allowing free \ce{^15N} to enhance the \ce{HC^15N} abundance. This fractionation process is enhanced in strongly irradiated regions of the disk, such as the disk atmosphere or cavities of transition disks. Evidence of an efficient illumination of the PDS~70 cavity was presented by \cite{law2024mapping}, along with its impact on line emission morphology of various molecular tracers \citep{rampinelli2024}. In this context, radial profiles of isotope fractionation can reveal how different fractionation processes dominate at different planet forming scales.

\begin{figure*}[ht!]
    \centering
    \includegraphics[scale=0.24]{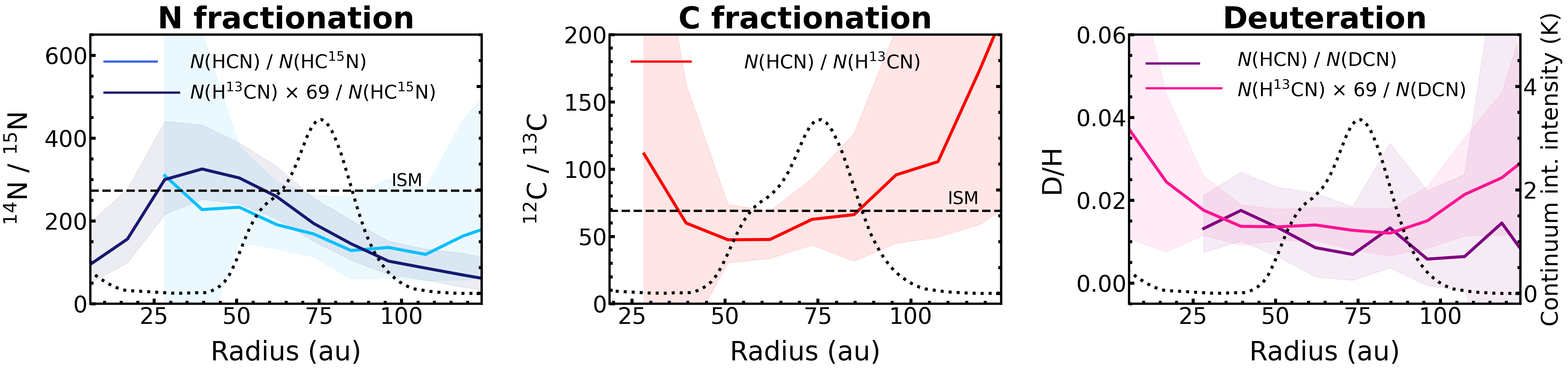}
    \caption{Nitrogen, carbon, and hydrogen fractionation of the HCN molecule. Left panel: $69\times\ce{H^13CN}/\ce{HC^15N}$ (dark blue) and \ce{HCN}/\ce{HC^15N} (light blue) profiles. Middle panel:  \ce{HCN}/\ce{H^13CN}. Right panel: $69\times\ce{H^13CN}/\ce{DCN}$ (pink) and \ce{HCN}/\ce{DCN} (purple) profiles. Ribbons show the 16th and 84th percentiles of the posterior distributions of isotopologue ratios. The dashed horizontal black lines show the ISM values of $\ce{^14N}/\ce{^15N}=274$ (blue) and of $\ce{^12C}/\ce{^13C}=69$, while the ISM values of D/H is not indicated, as it is $\sim10^{-5}$ \citep[see references in][]{nomura2022isotopic}. The marker length in each legend shows the major axis of the beam of cubes used to retrieve the corresponding profiles in the plots. The black dotted line shows the $855\mu$m continuum integrated intensity profile \citep{isella2019detection,benisty2021circumplanetary}.}
    \label{fig:12C13C}
\end{figure*}

In particular, radial profiles of \ce{^14N}/\ce{^15N} have been extracted for the disks around TW~Hya (from CN and HCN isotopologues, \citealt{hily-blant2017,hily2019multiple}) and V4046~Sgr (from HCN isotopologues, \citealt{nomura2022isotopic}, Guzmán et al. in prep.). The profiles are reconstructed up to $\sim$60~au, and they are both increasing with radius, as shown in the left panel of Fig~\ref{fig:frac_disks}. This has been interpreted as the result of isotope selective photodissociation: the higher gas density and stronger UV field at smaller radii result in an efficient self-shielding of \ce{N2}, and efficient photodissociation of \ce{N^15N}, thus resulting in enhanced \ce{HC^15N} in the inner region of the disks. These observational results also agree with fractionation models \citep{visser2018nitrogen, lee2021modeling}, and are consistent with the result we found for PDS~70 (see left panel in Fig.~\ref{fig:frac_disks}). However, we highlight that the spatial resolution and sensitivity make the profiles shown in Fig.~\ref{fig:N_D_frac}, \ref{fig:12C13C}, \ref{fig:frac_disks} highly uncertain for radii $\lesssim 25$~au, and therefore need to be interpreted with caution.

While radial profiles of HCN/\ce{HC^15N} for TW~Hya and V4046~Sgr are only probing the increasing trend inside $\sim 60$~au, we could extract the isotopologue ratio out to $\sim120$~au for the PDS~70 disk. The \ce{^14N}/\ce{^15N} ratio shows a turnover at $\sim60$~au, and it decreases below the ISM value at larger radii. This trend agrees with the predictions from the thermo-chemical models presented by \cite{visser2018nitrogen} for a typical disk around a T~Tauri star (full disk). As they show in their Fig.~12, the HCN/\ce{HC^15N} column density ratio is expected to increase up to $\sim 400$ out to $\sim30$~au and then decrease down to $\sim260$ between $\sim30$ and $100$~au. \cite{visser2018nitrogen} showed that both the inner increase and the outer decrease in the HCN/\ce{HC^15N} column density ratio are the result of isotope selective photodissociation of \ce{N2}, as low-temperature isotope exchange reactions alone do not produce any important radial variations. This qualitative agreement between the result directly extracted from ALMA observations in this work, and the prediction from thermo-chemical models obtained assuming a typical disk around a T-Tauri star \citep{visser2018nitrogen}, suggests that regardless of the physical structure of the individual source which is analyzed, isotope selective photodissociation plays a fundamental role in changing the HCN fractionation during the protoplanetary disk stage. This is also consistent with the fact that the decrease of HCN/\ce{HC^15N} in the outer disk is always predicted from thermo-chemical models by \cite{visser2018nitrogen}, even when varying disk parameters, such as the disk mass, flaring angle, large grain fraction, stellar type, or accretion rate.

The ratio of small to large grains has the largest quantitative impact on the isotopologue ratio \citep{visser2018nitrogen}, as it affects the attenuation of UV photons, which are responsible both for the production and fractionation of HCN in the gas phase. The presence of an inner disk in the PDS~70 system, observed both in IR and sub-mm observations, and the large mm-dust cavity ($\sim75$~au) can therefore have a role in setting the fractionation levels of HCN, and the radial location of the turn over in the HCN/\ce{HC^15N} profile (see the left panel in Fig.~\ref{fig:frac_disks}). In this context, both TW~Hya and V4046~Sgr also show a dust cavity of $\sim2.5$~au \citep{andrews2016ringed} and $\sim13$~au \citep{martinez2022high}, respectively, smaller with respect to the PDS~70 one, which is also showing a deep gas gap. The location of the layer where isotope selective photodissociation dominates is also related to the distribution of the $\mu$m dust, which attenuates the UV flux. However, our attempt of extracting the HCN emitting layer from the radial profile of its excitation temperature, to compare it to the $\mu$m dust surface, is inconclusive (see Appendix~\ref{appendix:layer_HCN}). Specific thermo-chemical models for the PDS~70 disk are therefore needed to quantitatively reproduce the observed radial profile of the HCN/\ce{HC^15N} ratio.

Eventually, we highlight that low-temperature isotope exchange reactions, even if less efficient than isotope selective photodissociation \citep{roueff2015isotopic,wirstrom2018,loison2019chemical}, cannot be completely ruled out at low layers over the midplane or in the cold outer disk, outside the CO snowline ($\sim70$~au for the PDS~70 disk as estimated by \citealt{law2024mapping}.)


\subsection{Carbon}\label{subsec:disc:C}
The nitrogen fractionation of HCN shown in Fig.~\ref{fig:N_D_frac} is extracted assuming a radially constant \ce{^12C}/\ce{^13C} equal to the ISM value, which makes the above interpretation only speculative, as the role of nitrogen and carbon fractionation in determining the decreasing trend cannot be robustly distinguished with current observations. Nevertheless, our attempt of leveraging the HCN hyperfine component covered by the analysed observations (see Appendix~\ref{appendix:C}) suggests that the HCN/\ce{H^13CN} ratio is slightly increasing between $\sim$40 and 100~au, but the profile is consistent with the ISM value within the errorbars. In this picture, the assumption of $\ce{^12C/^13C}=69$ to convert the \ce{H^13CN} into and HCN column density is justified, as the HCN/\ce{HC^15N} profile still shows a decreasing trend between $\sim$40 and 90~au, which is consistent with the $69\times\ce{H^13CN}/\ce{HC^15N}$ profile. 

\begin{figure*}[ht!]
    \centering
    \includegraphics[scale=0.25]{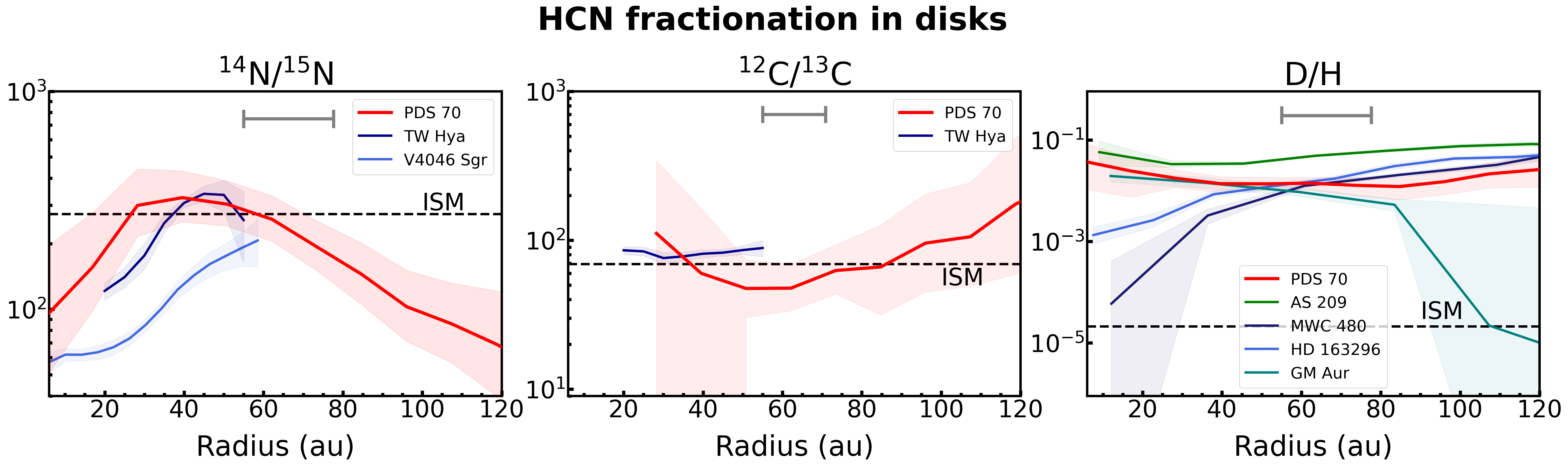}
    \caption{Radial profiles of nitrogen, carbon, and hydrogen fractionation of HCN in disks. The red profiles in the three panels show the results for the PDS~70 disk (solid lines and ribbons show the 50th, 16th, and 84th percentiles of the posterior distributions, respectively), while the horizontal dashed lines indicate the ISM values \citep{nomura2022isotopic, ritchey2015, wilson1999isotopes, linsky2006}. The isotopic ratios profiles for the PDS~70 disk are compared to the ones for the TW~Hya \citep{hily-blant2019multiple} and V4046~Sgr disks (\citealt{nomura2022isotopic}, Guzmán et al. in prep.) for nitrogen in the left panel, the TW~Hya disk \citep{hily-blant2019multiple} for carbon in the middle panel, and to the disks in the MAPS survey \citep{cataldi2021molecules} for hydrogen in the right panel. The nitrogen and hydrogen fractionation profiles are obtained by converting the \ce{H^13CN} into an HCN column density, assuming a fixed $\ce{^12C}/\ce{^13C}=69$ ratio. The grey line on top of each panel shows the major axis of the beam of the observations used to extract the profiles for the PDS~70 disk.}
    \label{fig:frac_disks}
\end{figure*}

To the best of our knowledge, the carbon isotopic ratio has been extracted only for one other disk, i.e. the disk around TW~Hya, for CO \citep{zhang2017mass,yoshida2022}, HCN \citep{hily-blant2019multiple}, CN \citep{yoshida2024}, and CCH \citep{bergin2024}. These results show that the value of \ce{^12C/^13C} can vary between $\sim20$ to $\sim90$ depending on the molecule used to extract it, and evidence for two separate carbon isotopic reservoirs \citep{bergin2024}. The \ce{^12C/^13C} ratio extracted from HCN in TW~Hya is shown in the middle panel of Fig.~\ref{fig:frac_disks}, with \ce{HCN}/\ce{H^13CN} being mostly constant and above the ISM value up to $\sim60$~au.

On the other hand, our result for the PDS~70 disk shows that carbon fractionation may become relevant beyond $\sim$90~au (see middle panel of Fig.~\ref{fig:12C13C}), which could be linked to the onset of the low-temperature isotope exchange reaction $\ce{^13C} + \ce{CO} \rightleftarrows \ce{C^+} + \ce{^13CO} + \mathrm{35}$~K in the cold outer disk, depleting \ce{^13C^+} in favor of \ce{^13CO} and increasing the HCN/\ce{H^13CN} ratio \citep[see e.g.][and references therein]{visser2018nitrogen, oberg2021astrochemistry}. This result qualitatively agrees also with predictions from fractionation models performed by \cite{visser2018nitrogen} for HCN/\ce{H^13CN}. This can also affect the result we presented for nitrogen and hydrogen fractionation in Fig.~\ref{fig:N_D_frac}, which was extracted assuming a radially constant \ce{^12C/^13C} ratio. The low values of the \ce{^14N/^15N} and \ce{H/D} ratios (dark blue and pink profiles in Fig.~\ref{fig:N_D_frac} respectively) outside $\sim 100$~au could be an artificial result of the assumption of a radially constant \ce{^12C/^13C}, instead of nitrogen or hydrogen fractionation in the outer disk.

However, the picture of carbon fractionation is complex: reactions producing HCN can start either from \ce{C^+} or \ce{C}, depending on the region in the disk \citep{visser2018nitrogen}. Therefore, both low temperature isotope exchange reactions and isotope selective photodissociation can play a crucial role in determining carbon fractionation of HCN. Moreover, these reactions primarily depend on the main C-carrier, i.e. CO, whose isotopic ratio should be inferred to better interpret the result for HCN. In this context, further thermochemical modeling specifically tuned on the transition disk of PDS~70, and higher spectral resolution observations fully covering also the main and hyperfine components of the HCN (4-3) transition are needed to interpret the result for HCN fractionation and robustly disentangle carbon and nitrogen fractionation. 

\subsection{Hydrogen} \label{subsec:disc:H}
As for nitrogen fractionation, we extracted the deuteration profile of HCN, either by $N(\ce{DCN})/[N(\ce{H^13CN})\times 69]$, or by $N(\ce{DCN})/N(\ce{HCN})$ (pink and purple line in the right panel of Fig.~\ref{fig:12C13C}, respectively), leveraging the resolved hyperfine component of the HCN line (see Appendix~\ref{appendix:C}). The deuteration profile is almost flat (around $\sim0.02$, for radii between 40 and 100~au. The increase at $\sim$100 au in the pink profile in the right panel of Fig.~\ref{fig:12C13C} could indicate a more efficient deuteration pathway in the cold outer disk, but the profile may be affected by carbon fractionation. The \ce{^12C}/\ce{^13C} profile shown in Fig.~\ref{fig:12C13C}\sout{)} has large uncertainties outside $\sim90$~au, and thus does not allow to robustly conclude on the contribution of carbon fractionation in the outer disk. Similarly to nitrogen fractionation, in order to definitely disentangle the effect of carbon fractionation, new high spectral resolution observations covering the two hyperfine components of the HCN line are needed.

The deuteration level of the HCN molecule in the PDS~70 disk is significantly above the ISM value of $\sim 10^{-5}$ \citep{linsky2006, nomura2022isotopic}, which suggests that there are efficient deuteration processes enriching HCN with deuterium.
The main deuteration pathway in protoplanetary disks is through isotope exchange reactions involving HD at low temperatures ($T\lesssim25$~K) or \ce{CH3^+} at higher temperatures ($T\lesssim300$~K) \citep{millar1989deuterium,aikawa1999,stark1999}, which produce \ce{H2D^+} and \ce{CH2D^+}, respectively. Deuterium from \ce{H2D^+} and \ce{CH2D^+} can be transferred to other molecules through proton transfer such as HNC + \ce{H2D^+} $\rightarrow$ \ce{DCNH^+} + \ce{H2}, which can subsequently produce DCN through dissociative recombination \ce{DCNH^+} + \ce{e^-} $\rightarrow$ DCN + H \citep{cataldi2021molecules}.

Given the different gas temperatures at which these reactions set, we expect them to dominate in different regions in the disk. The radial profile of DCN/HCN ratio is mostly constant at $\sim0.02$ between $\sim40$ and 100~au, which suggests that both reactions are important in setting the deuteration level of HCN, even if at different radial scales.

\begin{figure*}[ht!]
    \centering
    \includegraphics[scale=0.45]{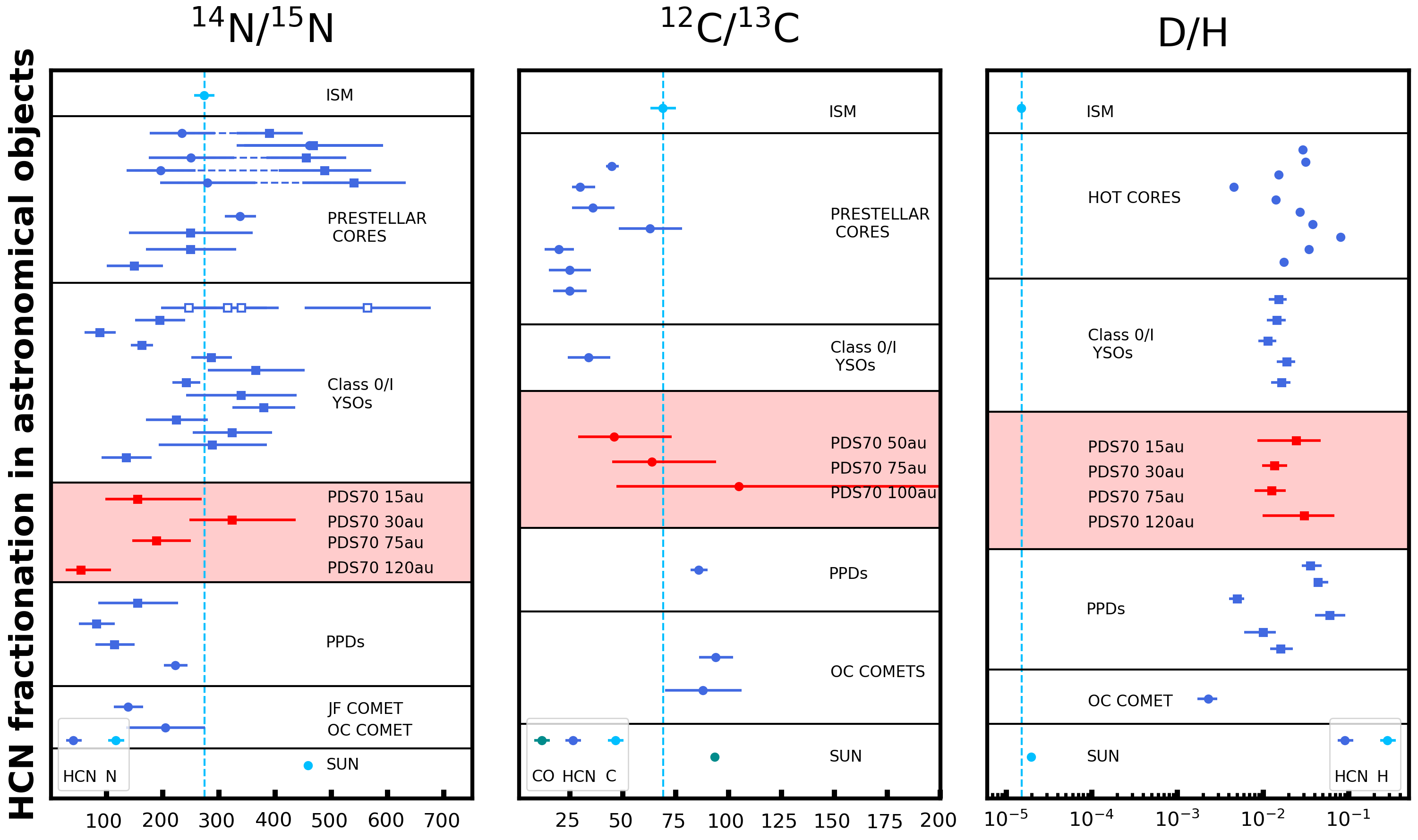}
    \caption{Nitrogen, carbon, and hydrogen fractionation in various astronomical objects. Red and dark blue markers indicate isotope ratios of HCN, with red dots referring to values extracted at different radii for the PDS~70 disk in this work. Light blue markers show isotopic ratios. The green marker in the middle panel for the sun is obtained from CO. The dashed light blue lines refer to the ISM isotopic ratios \citep{linsky2006, wilson1999isotopes,ritchey2015}. Errorbars are not displayed if the uncertainty is too low to be visible in the plot. Squared markers indicate indirect measurements of DCN/HCN and HCN/\ce{HC^15N} ratios, assuming a fixed \ce{^12C/^13C}, while circles refer to results obtained directly from the main isotopologue HCN. See Appendix~\ref{appendix:other_stages} for references on the values presented here, and related methods. For a few sources both direct and indirect estimates are available and displayed on the same row connected by a dashed horizontal lines. Aligned empty squares refer to ratios extracted for the same source but in different regions.}
    \label{fig:frac_stages}
\end{figure*}

The right panel of Fig.~\ref{fig:frac_disks} shows the comparison between the radial profile of HCN deuteration in the PDS~70 disk (red profile) and the MAPS survey \citep{cataldi2021molecules}. 
The D/H values show a large scatter among different sources, but they are systematically above the elemental ISM value (dashed black horizontal line).
This suggests that in situ fractionation processes are actively resetting the original ISM value in a source-dependent way, which is strongly affected by different physical conditions set in different disks, such as gas temperature or ionization structure. In particular, as resulting from the MAPS survey \citep{cataldi2021molecules}, the DCN/HCN profile in disks around Herbig stars show a steeper decreasing profile of DCN/HCN at small radii with respect to T~Tauri disks (see the right panel of Fig.~\ref{fig:frac_disks}). This has been interpreted as the effect of an higher temperature at small radii for Herbig with respect to T~Tauri disks, thus inhibiting the deuteration reaction. This picture is consistent with the fact that the radial profile of the DCN/HCN ratio inside $\sim100$~au for the T~Tauri disk of PDS~70 is compatible with the ones for T~Tauri disks in the MAPS sample.

\subsection{Comparison with early and later stages}\label{subsec:comparison_objects}

Fig.~\ref{fig:frac_stages} shows a summary of nitrogen, carbon, and hydrogen fractionation in various astronomical objects across different evolutionary stages of star and planet formation from the literature. The red and dark blue markers indicate isotopic ratios of the HCN molecule for the PDS~70 disk (at different radii) and in other sources (prestellar/protostellar cores, class 0/I sources, comets), respectively. Light blue markers refer to isotopic ratios measured in the ISM and the sun, with the exception of the solar \ce{^12C/^13C} measured from CO (green marker in the middle panel). Appendix~\ref{appendix:other_stages} lists all the references for the values presented in Fig.~\ref{fig:frac_stages} and some details on the adopted methodologies.

The isotopologue ratios of HCN across early stages of the star formation history, i.e. in prestellar cores and class 0/I Young Stellar Objects (YSOs), are in most of the cases consistent with the results presented in this work for the PDS~70 disk. This is suggesting either a conservation of the original gas phase isotopologue fractionation, or that if gas phase fractionation processes are fast and the physical conditions in each stage are similar, the resulting isotopologue ratios are consistent. On the other hand, both HCN/\ce{HC^15N} and DCN/HCN ratios in prestellar/protostellar cores show a large spread, possibly highlighting the strong dependence of fractionation processes on the local physical condition in different sources, such as the gas temperature for isotope exchange reactions and radiation field for isotope selective photodissociation. Moreover, HCN/\ce{HC^15N} values extracted for early stages presented in Fig.~\ref{fig:frac_stages} include both direct and indirect methods relying on various assumptions such as the \ce{^12C/^13C} conversion factor, or the lines opacities \citep[][see also Appendix~\ref{appendix:other_stages}]{hily-blant2020,jensen2024fractionation}. \cite{hily-blant2020} highlighted how indirect methods result in more scattered \ce{^14N/^15N} values compared to direct methods. In particular, the first direct estimates of HCN/\ce{H^13CN} and HCN/\ce{HC^15N} for a prestellar core were presented by \cite{magalhaes2018nitrogen}, showing deviations from indirect methods pointing to a \ce{^13C} enrichment. Similarly, \cite{jensen2024fractionation} presented both direct and indirect measurements of the HCN/\ce{HC^15N} ratio towards six starless and prestellar cores, showing significant difference (as also visible from Fig.~\ref{fig:frac_stages}). Consistent with what we previously highlighted, \cite{jensen2024fractionation} also showed that there is no evident trend among different stages of star formation when considering direct estimates. It is also important to highlight that in the YSOs phase there could be a gas-phase contribution of HCN directly released from icy grains, even if gas-phase formation of HCN is thought to be dominant \citep{bergner2020deuteration}. 
The gas-phase origin of HCN in YSOs is supported by single dish observations where the HCN sublimation region (HCN desorption temperature of $\sim3370$~K, \citealt{noble2012desorption},\citealt{bergner2022HCN}) is unresolved \citep{cataldi2021molecules}. Finally, as shown in Fig.~\ref{fig:frac_stages}, \cite{evans2022fractionation} presented the HCN/\ce{HC^15N} ratio measured in different regions of the protocluster OMC-2 FIR4, showing consistent results for the central region and a higher ratio for the East region, which is more distant from the protocluster center. This highlights again the importance of studying spatial variations of fractionation levels in the same source, since fractionation processes are strongly affected by the local physical conditions.

A spread of \ce{HCN/HC^15N} and DCN/HCN ratios is also observed in protoplanetary disks (PPDs) \citep{huang2017alma,guzman2017,hily-blant2019multiple}, with HCN/\ce{HC^15N} values in agreement with measurements both for early and later stages. DCN/HCN ratios show approximately an order of magnitude difference between disks around T~Tauri and Herbig stars \citep{huang2017alma}, with the lowest deuteration levels measured in Herbig disks, possibly due to their higher temperature with respect to T~Tauri disks, as already mentioned in Sect.~\ref{subsec:disc:H}. The T~Tauri disk of V4046~Sgr is an exception in this picture, with its low value of $\ce{DCN/HCN}\sim50$, which is also only a factor $\sim2$ higher than the DCN/HCN ratio measured for the Hale-Bopp comet \citep{meier1998deuteration}. In this context, we highlight that V4046~Sgr is the oldest disk in the sample, being $\sim 23$~Myr old \citep{mamajek2014}, which could also affect its deuteration level. Moreover, we highlight that HCN isotopologue ratios presented in Fig.~\ref{fig:frac_stages} for protoplanetary disks are only disk integrated, while radial variations presented in Fig.~\ref{fig:frac_disks} reveal how the local environment affects fractionation processes.

HCN/\ce{H^13CN} ratios seem to suggest an increasing evolutionary trend, with ratios in early stages and comets matching the HCN/\ce{H^13CN} values at small and large radii in the PDS~70 disk, respectively. On the other hand, the large uncertainties on the HCN/\ce{H^13CN} radial profile in the PDS~70 disk do not allow to assess the robustness of this trend. Both richer statistics on carbon fractionation for early and later stages of star and planet formation and higher spectral resolution observations for the PDS~70 disk could shed light on the origin of this possible trend.

Moving to the comparison with later stages, we compare the HCN isotopologue ratios extracted for the PDS~70 disk with cometary values, both for Jupiter Family (JF) and Oort Family (OF). The \ce{^12C/^13C} and \ce{^14N/^15N} ratios are in good agreement with the values we found in the PDS~70 disk, especially with the outer disk region outside $\sim70$~au, which seems to suggest a correlation between the cometary material and the gas phase composition in the outer disk \citep{bockelee2008HCNcomets, cordiner2019carbon}.
On the other hand, this picture is not consistent with the only measurement of DCN/HCN in the Hale-Bopp comet which is approximately one order of magnitude lower than the PDS~70 disk values \citep{meier1998deuteration}. Moreover, both the DCN/HCN and HCN/\ce{H^13CN} ratios in comets are not consistent with the corresponding ratios measured in prestellar/protostellar cores.

The low value of DCN/HCN in Hale-Bopp is suggesting that the HCN ice observed in the comet may not be probing the same gas phase HCN reservoir observed in prestellar/protostellar cores. In particular, the ice component could originate from the early phases of star and planet formation. Due to the low formation efficiency of HCN ice, it is undetected in the ISM though a low abundance is expected \citep{gerakines2004ultraviolet, mumma2011, gerakines2022direct,mcclure2023ice}. The cometary HCN deuteration could have also been altered once HCN was already incorporated on the ice. In particular, thermally- or radiation-induced exchange reactions on the ices were shown to play an important role in altering the deuteration of a variety of ices, such as hydrocarbons- or methanol-water ices \citep{mousis200deuterium,weber2009hydrogen,faure2015methanol}.

In summary, we highlight that this is still a very uncertain picture, requiring further investigation. In particular, better statistics on the DCN/HCN in comets could help understanding if the value measured for the Hale-Bopp comet is an outlier or typical for comets. HCN fractionation is generally poorly constrained in comets, thus making it difficult to speculate on the origin. Similarly, it is important to probe HCN carbon fractionation in more objects throughout the star and planet formation timeline, to assess the suggested evolutionary trend. Moreover, further theoretical modeling and laboratory studies on HCN ices are needed to shed light on the role of deuteration in the ice phase.

Finally, we highlight that we could not directly compare our results with the recent measurements of isotopologue ratios in exoplanetary atmospheres \citep{zhang202113co,gandhi2023}, as the latter are not obtained from the same tracers (\ce{NH3} and CO) as the former (HCN), and could therefore be driven by very different fractionation processes from the ones we are probing with observations used in this work.

\section{Conclusion} \label{sec:conclusion}

In this work we present radial profiles of carbon, nitrogen, and hydrogen fractionation of the HCN molecule in the planet-hosting disk around PDS~70, showing how local fractionation pathways can alter isotopologue ratios. We summarize the main results as follow:
\begin{itemize}
    \item We show spatially resolved emission of HCN ($4-3$), \ce{H^13CN} ($3-2$), \ce{H^13CN} ($4-3$), \ce{HC^15N} ($3-2$), \ce{HC^15N} ($4-3$), and DCN ($3-2$) in the PDS~70 disk.
    \item We present a decreasing trend of the \ce{^14N}/\ce{^15N} isotopic ratio of HCN, outside the disk cavity wall, which we suggest to be linked to isotope selective photodissociation of \ce{N2}.
    \item We show that the D/H profile of HCN is almost flat throughout the disk extent and slightly decreasing in the outer disk. The deuteration profile does not show any correlation with the nitrogen fractionation one, which is consistent with the fact that deuteration is thought to be mainly driven by gas-phase isotope-exchange reactions in the cold midplane/outer disk.
    \item We retrieve the radial profile of the  \ce{HCN}/\ce{H^13CN} ratio, by extracting the HCN column density leveraging the spatially and spectrally resolved hyperfine component of HCN (4-3). The extracted \ce{^12C}/\ce{^13C} ratio is consistent with the ISM value. This suggests that the decreasing trend in \ce{H^13CN}/\ce{HC^15N} is primarily driven by local nitrogen fractionation pathways.
\end{itemize}

Radial variations of the gas-phase isotopologue ratios reveal local fractionation pathways in the PDS~70 disk, which could leave an imprint on the gas accreted by the two forming planets. This highlights the potential of isotopic ratios to reconstruct the journey of planetary material. Higher spectral resolution observations, also covering the high velocity hyperfine component of the HCN line are needed to robustly confirm our results. Moreover, comparison with thermochemical models specifically tuned on PDS~70 could help interpret observations in light of possible fractionation pathways, and in particular better constrain the role of carbon fractionation in the main C-carrier, i.e. CO.

This work adds to the list of the few protoplanetary disks with a detailed analysis of their fractionation profiles, with PDS 70 being the only target where the photospheres of embedded protoplanets can be observed and characterized. Even though modelling is needed to connect the fractionation of different molecular or atomic species that best trace individual evolutionary steps in the star and planet formation process, promoting a direct comparison of isotopologue ratios in exoplanetary atmospheres and protoplanetary disks is of primary importance to build a bridge between formed planets and their natal environment.

\begin{acknowledgements}

We thank V.V. Guzmán, P. Hily-Blant, and G. Cataldi for their availability in sharing results particularly important for the purpose of this work. 

This paper makes use of the following ALMA data: 

ADS/JAO.ALMA\#2019.1.01619.S,

ADS/JAO.ALMA\#2022.1.01695.S. 

ALMA is a partnership of ESO (representing its member states), NSF (USA) and NINS (Japan), together with NRC (Canada), MOST and ASIAA (Taiwan), and KASI (Republic of Korea), in cooperation with the Republic of Chile. The Joint ALMA Observatory is operated by ESO, AUI/NRAO and NAOJ.

L.R., S.F., and M.L. are funded by the European Union (ERC, UNVEIL, 101076613). Views and opinions expressed are however those of the authors only and do not necessarily reflect those of the European Union or the European Research Council. Neither the European Union nor the granting authority can be held responsible for them. S.F. also acknowledges financial contribution from PRIN-MUR 2022YP5ACE.

P.C. acknowledges support by the ANID BASAL project FB210003.

M.B. has received funding from the European Research Council (ERC) under the European Union’s Horizon 2020 research and innovation programme (PROTOPLANETS, grant agreement No. 101002188).

Support for C.J.L. was provided by NASA through the NASA Hubble Fellowship grant No. HST-HF2-51535.001-A awarded by the Space Telescope Science Institute, which is operated by the Association of Universities for Research in Astronomy, Inc., for NASA, under contract NAS5-26555.

B.P.R. is supported by NASA STScI grant JWST-GO-01759.002-A. The Center for Exoplanets and Habitable Worlds is supported by the Pennsylvania State University and the Eberly College of Science.

The PI acknowledges the use of
computing resources from the Italian node of the European ALMA Regional
Center, hosted by INAF-Istituto di Radioastronomia.
\end{acknowledgements}

\bibliographystyle{aa}
\bibliography{references}

\onecolumn
\begin{appendix}

\section{Fluxes of Band 6 and 7 molecular lines}\label{appendix:fluxes}

The spectral set up of Band~7 data (\#2022.1.01695.S, PI M. Benisty) includes other molecular lines in addition to the HCN isotopologues ones: \ce{^12CO} (3-2), CS (7-6), \ce{HCO^+} (4-3), and SO ($8_8-7_7$). Rest frequencies, imaging parameters, and line fluxes of the detected molecular transitions are listed in Table~\ref{tab:lines_b6b7}. Cubes were obtained using natural weighting. Line fluxes are obtained by spatially integrating the integrated intensity maps of each line inside a de-projected circle with radius 3$\as$ for the most extended \ce{^12CO}, and 2$\as$ for the others.  The uncertainty is evaluated as the standard deviation of the flux measured in 26 de-projected circles, outside the emitting area of the integrated intensity maps, with the same radius used to extract the flux. We took the maximum number of circles that was possible to place on the image without overlap, and used not primary beam corrected images, to ensure uniform noise \citep{rampinelli2024}. Table~\ref{tab:lines_b6b7} lists also detected lines in Band~6 data (\#2019.1.01619.S, PI S. Facchini): the reader is referred to \cite{rampinelli2024} for more details on imaging strategies and line flux evaluation.

\begin{table*}
  \renewcommand{\arraystretch}{1.2}
  \centering
  \caption{Detected lines in Band~6 and 7 observations, rest frequencies, imaging properties (channel width, beam, RMS, and fluxes).}
  \label{tab:lines_b6b7}
  \begin{threeparttable}
  \begin{tabular}{ c c c c c c c }
      \hline
      \hline
      Species & Transition\tnote{a} & Frequency &  $\Delta \mathrm{v}$ & Beam & RMS & Flux\tnote{b} \\
      &  & [GHz] & [km~s$^{-1}$] &  & [mJy~beam$^{-1}$] & [mJy km~s$^{-1}$]\\
      \hline
      $^{12}$CO & 2-1 & 230.5380 & 0.1 & $0\farcs28 \times 0\farcs22$ & 0.54 & 6175 $\pm$ 23 \\
      $^{13}$CO & 2-1 & 220.3986 & 0.2 & $0\farcs30 \times 0\farcs24$ & 0.42 & 1859 $\pm$ 12 \\
      C$^{18}$O & 2-1 & 219.5603 & 0.2 & $0\farcs31 \times 0.24$ & 0.33 & 466 $\pm$ 7 \\
      H$_2$CO & 3$_{0,3}$-2$_{0,2}$ & 218.2221 & 0.4 & $0\farcs29 \times 0\farcs23$ & 0.19 & 708 $\pm$ 7 \\
      H$_2$CO & 3$_{2,1}$-2$_{2,0}$ & 218.7600 & 0.4 & $0\farcs29 \times 0\farcs23$ & 0.20 & 57 $\pm$ 6 \\
      C$_2$H & $3_{7/2, 4}$-$2_{5/2, 3}$ & 262.0042 & 0.4 & $0\farcs20 \times 0\farcs15$ & 0.16 & 723 $\pm$ 10 \\
      C$_2$H & $3_{7/2, 3}$-$2_{5/2, 2}$ & 262.0064 & 0.4 & $0\farcs20 \times 0\farcs15$ & 0.16 & 577 $\pm$ 10 \\
      H$^{13}$CN & 3-2 & 259.0117 & 0.4 & $0\farcs20 \times 0\farcs15$ & 0.13 & 360 $\pm$ 8 \\
      HC$^{15}$N & 3-2 & 258.1571 & 0.4 & $0\farcs20 \times 0\farcs15$ & 0.15 & 176 $\pm$ 7 \\
      DCN & 3-2 & 217.2386 & 0.4 & $0\farcs29 \times 0\farcs23$ & 0.22 & 230 $\pm$ 6 \\
      H$^{13}$CO$^+$ & 3-2 & 260.2553 & 0.4 & $0\farcs21 \times 0\farcs15$ & 0.16 & 372 $\pm$ 7 \\
      CS & 5-4 & 244.9355 & 1.4 & $0\farcs22 \times 0\farcs15$ & 0.06 & 445 $\pm$ 6 \\
      \ce{c-C3H2} & 3$_{2,1}$-2$_{1,2}$ & 244.2221 & 1.4 & $0\farcs22 \times 0\farcs16$ & 0.07 & 16 $\pm$ 2 \\
      \ce{c-C3H2} & 7$_{1,6}$-7$_{0,7}$ & 218.7327 & 0.4 & $0\farcs29 \times 0\farcs23$ & 0.20 & 10 $\pm$ 3\tnote{c} \\
      \ce{c-C3H2} & 7$_{2,6}$-7$_{1,7}$ & 218.7327 & 0.4 & $0\farcs29 \times 0\farcs23$ & 0.69 & 10 $\pm$ 3\tnote{c} \\
      \hline
      $^{12}$CO & 3-2 & 345.7960 & 0.9 & $0\farcs13 \times 0\farcs12$ & 0.76 & 13632 $\pm$ 42 \\ 
      CS & 7-6 & 342.8829 & 0.9 & $0\farcs13 \times 0\farcs12$ & 0.80 & 407 $\pm$ 31 \\ 
      \ce{HCO^+} & 4-3 & 356.7342 & 0.9 & $0\farcs13 \times 0\farcs11$ & 0.96 & 5470 $\pm$ 31 \\
      \ce{H^13CN} & 4-3 & 345.3398 & 0.9 & $0\farcs13 \times 0\farcs11$ & 0.70 & 530 $\pm$ 21 \\
      \ce{HC^15N} & 4-3 & 344.2003 & 0.9 & $0\farcs13 \times 0\farcs12$ & 0.75 & 234 $\pm$ 34 \\
      \ce{HCN}\tnote{d} & 4-3 & 354.5055 & 0.9 & $0\farcs37 \times 0\farcs34$ & 1.82 & 5902 $\pm$ 45 \\
      \hline
      \hline
  \end{tabular}
  \begin{tablenotes}
      \centering
      \item[a] Quantum numbers are formatted for \ce{H2CO} \citep[CDMS, ][]{chardon1973structure, muller2017submillimeter}, \ce{c-C3H2} \citep[CDMS, ][]{bogey1986centrifugal, vrtilek1987laboratory, lovas1992microwave} as $N_{K_a, K_c}$, for C$_2$H as $N_{J, F_1}$ \citep[CDMS, ][]{sastry1981laboratory, muller2000submillimeter, padovani2009c2h}, for \ce{^12CO} \citep[CDMS, ][]{winnewisser1997sub}, \ce{^13CO} \citep[CDMS, ][]{klapper2000sub, cazzoli2004precise}, \ce{C^18O} \citep[CDMS, ][]{winnewisser1985millimeter}, \ce{H^13CN} \citep[CDMS, ][]{fuchs2004high, cazzoli2005lamb, maiwald2000pure}, \ce{HC^15N} \citep[CDMS, ][]{fuchs2004high, cazzoli2005lamb}, \ce{DCN} \citep[CDMS, ][]{brunken2004sub}, \ce{H^13CO^+} \citep[CDMS, ][]{gregersen2001erratum, lattanzi2007rotational}, \ce{CS} \citep[CDMS, ][]{muller2005cologne, bogey1981millimeter, ahrens1999pure}, \ce{HCO^+} \citep[CDMS, ][]{tinti2007hcop,lattanzi2007hcop} as $J$.
      \item[b] The uncertainty does not include the 10\% absolute flux calibration.
      \item[c] The value refers to the total flux of the blended \ce{c-C3H2} (218~GHz) lines.
      \item[d] Listed beam size, RMS, and flux are obtained from SB observations only.
  \end{tablenotes}
  \end{threeparttable}
\end{table*}

\FloatBarrier

\section{Imaging procedure}\label{appendix:imaging}

To retrieve the radial profiles of nitrogen, carbon, and hydrogen fractionation of HCN, we performed two rounds of imaging to match the spatial resolution of Band~6 and 7 observations, respectively. We chose the robust and/or Gaussian $uv$-taper which resulted in cubes with beam areas as close as possible in the two cases. The imaging parameters used in the two cases, resulting beam sizes and areas, and RMS are listed in Tab.~\ref{tab:imaging_molecules_2}. For the first case, only \ce{H^13CN}, \ce{HC^15N}, and DCN were imaged for the purpose of the analysis, while in the second case we also imaged the combined SB and LB observations of HCN. The resulting cube, with the corresponding beam size, beam area, and RMS listed in Tab.~\ref{tab:imaging_molecules_2} only covers half of the HCN (4-3) line.

\begin{table*}
  \centering 
  \caption{Weighting, $uv$-taper, beam size, beam area, epsilon values, and RMS of imaged cubes used in the analysis.}
  \label{tab:imaging_molecules_2}
      \begin{tabular}{c c c c c c c}
          \hline
          \hline
          Transition & robust & $uv$-taper & Beam & Beam area & $\epsilon$ & RMS \\
          & & & & [arcseconds$^2$] & & [mJy beam$^{-1}$] \\
          \hline
          H$^{13}$CN (3-2) & 0.67 & -- & $0\farcs15 \times 0\farcs11$ & 0.0178 & 0.45 & 0.60 \\
          H$^{13}$CN (4-3) & 2.00 & -- & $0\farcs13 \times 0\farcs12$ & 0.018 & 0.24 & 0.70 \\
          HC$^{15}$N (3-2) & 0.67 & -- & $0\farcs15 \times 0\farcs11$ & 0.018 & 0.45 & 0.71 \\
          HC$^{15}$N (4-3) & 2.00 & -- & $0\farcs13 \times 0\farcs12$ & 0.0178 & 0.24 & 0.75 \\
          DCN (3-2) & 0.30 & -- & $0\farcs14 \times 0\farcs11$ & 0.018 & 0.6 & 0.94 \\
          HCN (4-3) & 2.00 & -- & $0\farcs13 \times 0\farcs11$ & 0.017 & -- & 0.86 \\
          \hline
          H$^{13}$CN (3-2) & 2.00 & --& $0\farcs20 \times 0\farcs15$ & 0.035 & 0.22 & 0.55 \\
          H$^{13}$CN (4-3) & 2.00 & $0\farcs064$ & $0\farcs19 \times 0\farcs17$ & 0.035 & 0.37 & 0.75 \\
          HC$^{15}$N (3-2) & 2.00 & -- & $0\farcs20 \times 0\farcs15$ & 0.035 & 0.22 & 0.65 \\
          HC$^{15}$N (4-3) & 2.00 & $0\farcs064$ & $0\farcs19 \times 0\farcs17$ & 0.036 & 0.37 & 0.79 \\
          DCN (3-2) & 0.80 & -- &  $0\farcs19 \times 0\farcs15$ & 0.033 & 0.36 & 0.78 \\          
          \hline
          \hline
      \end{tabular}
\end{table*} 

\FloatBarrier

\section{Spectral smearing effect on brightness temperature}\label{appendix:B}
\begin{figure*}[ht!]
\centering
    \includegraphics[scale=0.5]{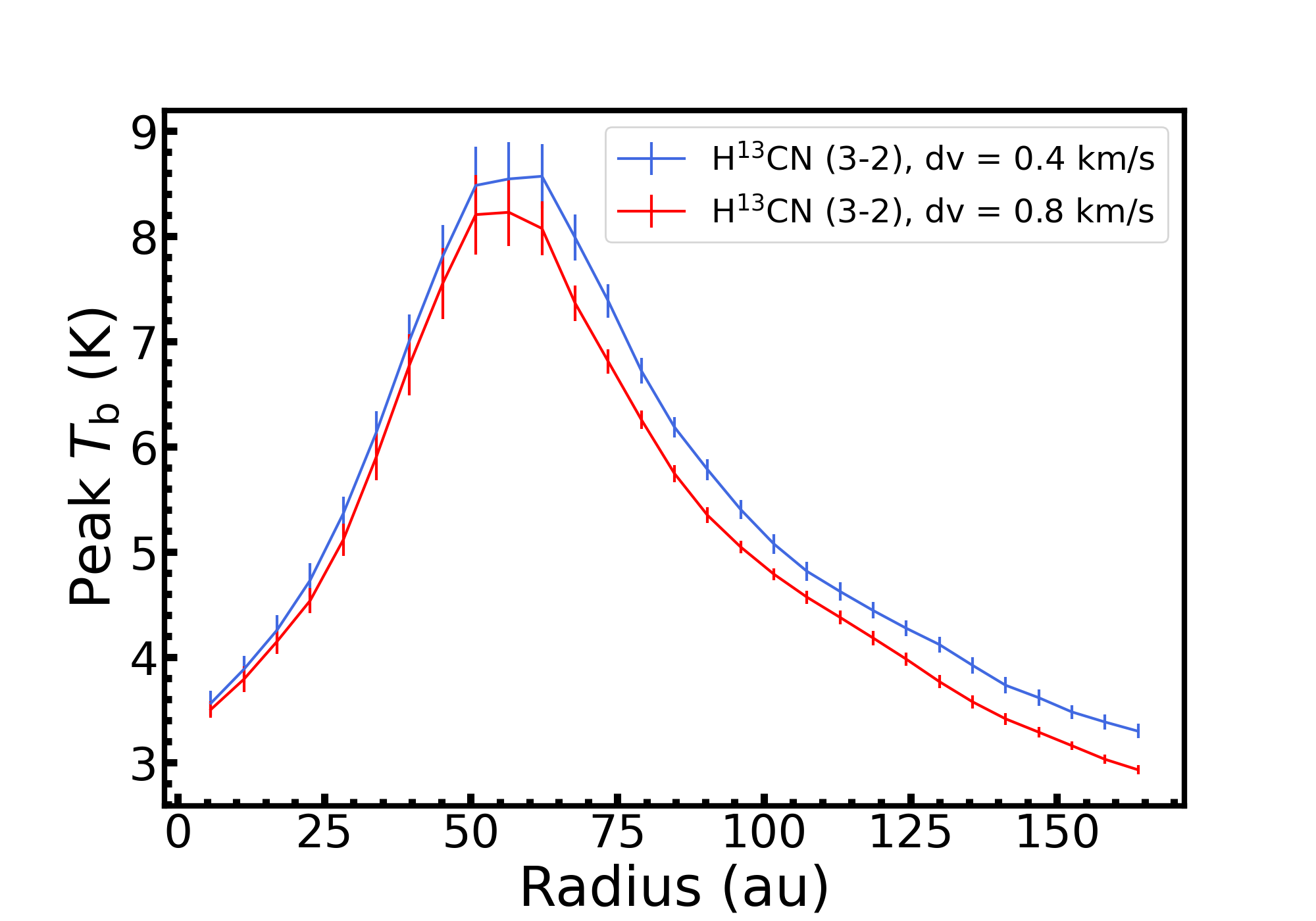}
    \caption{Radial profiles of peak brightness temperature extracted from cubes of the same \ce{H^13CN} (3-2) line, changing only the channel width in the imaging procedure (0.4~km~s$^{-1}$ blue line, 0.8~km~s$^{-1}$ red line). Error bars refer to the standard deviation of the brightness temperature in each annulus considered in the profile.}
    \label{fig:test_specsmearing}
\end{figure*}

In this work we evaluated the optical depth of the analysed lines starting from their peak brightness temperature and excitation temperature (see Eq.~\ref{eq:tau} in Sect.~\ref{sec:column}). However, the brightness temperature may be underestimated due to spectral smearing related to a finite spectral resolution. To quantitatively estimate the effect of spectral smearing we imaged the same cube of the \ce{H^13CN} (3-2) line twice, with a different channel width, and we extracted the radial profile of the peak brightness temperature, as shown in Fig.~\ref{fig:test_specsmearing}. We compared the profiles starting from a cube with a channel width of 0.4~km~s$^{-1}$ (blue profile), which is the smallest channel width allowed by the spectral resolution, and of 0.8~km~s$^{-1}$ (red profile).  As visible from Fig.~\ref{fig:test_specsmearing}, a lower peak brightness temperature corresponds to a larger channel width. Moreover, spectral smearing dominates at larger radii, which is expected by the decreasing line width. At smaller radii the effect is smaller, and also negligible with respect to statistical uncertainties. We highlight that even at larger radii a 10\% variation in the peak brightness temperature corresponds to roughly 1\% variation in the upper level population, for an excitation temperature of $\sim$30~K. This effect is thus negligible in the rotational diagram analysis we performed in this work.

\FloatBarrier

\section{\texorpdfstring{$^{12}$C/$^{13}$C}{12C/13C} ratio from the HCN hyperfine component} \label{appendix:C}

The \ce{^14N}/\ce{^15N} ratio of the HCN molecule in disks is typically extracted from \ce{H^13CN} and \ce{HC^15N}, assuming a fixed \ce{^12C}/\ce{^13C} ratio to convert the \ce{H^13CN} column density into an HCN column density, as the main isotopologue HCN is generally optically thick \citep[see e.g.][]{guzman2017}. On the other hand, when hyperfine components of HCN are spectrally resolved, the HCN column density, optical depth and excitation temperature can be inferred, together with their radial profiles if hyperfine components are also spatially resolved \citep[see e.g.][]{hily2019multiple, guzman2021molecules}. Long baseline observations of the HCN (4-3) line in the data we analysed only cover half of the main component spectrum, but they allow to resolve the hyperfine component at 354.5075~GHz. Figure~\ref{fig:hyperfine_fit} shows the HCN spectra (dark blue line) extracted averaging over annuli at different radii (indicated on the top right of each panel), corrected for the Keplerian rotation of the disk using the \texttt{GoFish} package \citep{teague2019gofish}. Only half of the main component is covered ($v_\mathrm{sys}=5.5$~km~s$^{-1}$), but the hyperfine component at 354.5075~GHz is marginally spectrally resolved (peak at $v=3.8$~km~s$^{-1}$), even with a spectral resolution of $\sim0.9$~km~s$^{-1}$. We extracted the flux associated with the hyperfine component by fitting a double Gaussian to the spectrum, stopping at 5.5~~km~s$^{-1}$. We fixed the center of the two Gaussian components to 3.8~km~s$^{-1}$ for the hyperfine and to 5.5~km~s$^{-1}$ for the main component respectively, and fit for the two widths and peaks. We sampled the posterior distribution using an MCMC sampler through the \texttt{emcee} package \citep{foreman2013emcee}. In order to take into account the spectral response, particularly critical for low spectral resolution, we spectrally over-sampled the Gaussian profiles and took the average value inside a spectral bin, to fit the observed spectrum. We applied the MCMC sampling to each annulus in which we divided the disk, with 128 walkers, 500 burn-in steps, and 500 steps, assuming a uniform prior of (0,5)~km~s$^{-1}$ for the two widths, of (0,10)~mJy~beam$^{-1}$ for the peak of the hyperfine component, and of (0,30)~mJy~beam$^{-1}$ for the peak of the main component.

Figure~\ref{fig:hyperfine_fit} shows the result of the MCMC sampling, for each annulus (the radius is indicated on the top right of each panel). The dashed light blue line is the sum of the two best fit Gaussian profiles, the dashed orange line shows the best fit Gaussian for the hyperfine component, the solid orange lines are 100 random samples of the posterior distribution, while the dark blue profile is the observed spectrum with the related uncertainty. The uncertainty for each spectrum was evaluated as standard deviation of the spectrum in a signal-free spectral range. We extracted the flux by integrating the Gaussian profiles of the hyperfine component obtained from the posterior distribution, with the best fit value being the 50th percentile of the posterior distribution of such fluxes, and the related uncertainty evaluated from the 16th and the 84th percentiles. From the flux of the hyperfine component, and the related uncertainty, we extracted the radial profile of the column density of HCN, assuming that the hyperfine component is optically thin, and the same excitation temperature inferred from the rotational diagram analysis of \ce{H^13CN} and \ce{HC^15N}. The fiducial radial range is between $\sim50$ and 120~au, because for small radii the two components are blended, and the fit is poorly constrained, while for large radii the fit converges towards larger line widths, which are not physical. On the other hand, in the fiducial range the Gaussian fit to the hyperfine component is consistent with a radially constant line width. In this radial range, we extracted the \ce{^12C}/\ce{^13C} profile shown in the middle panel of fig.~\ref{fig:12C13C}, taking the ratio between the HCN and the \ce{H^13CN} column densities.

\begin{figure*}[ht!]
    \centering
    \includegraphics[scale=0.16]{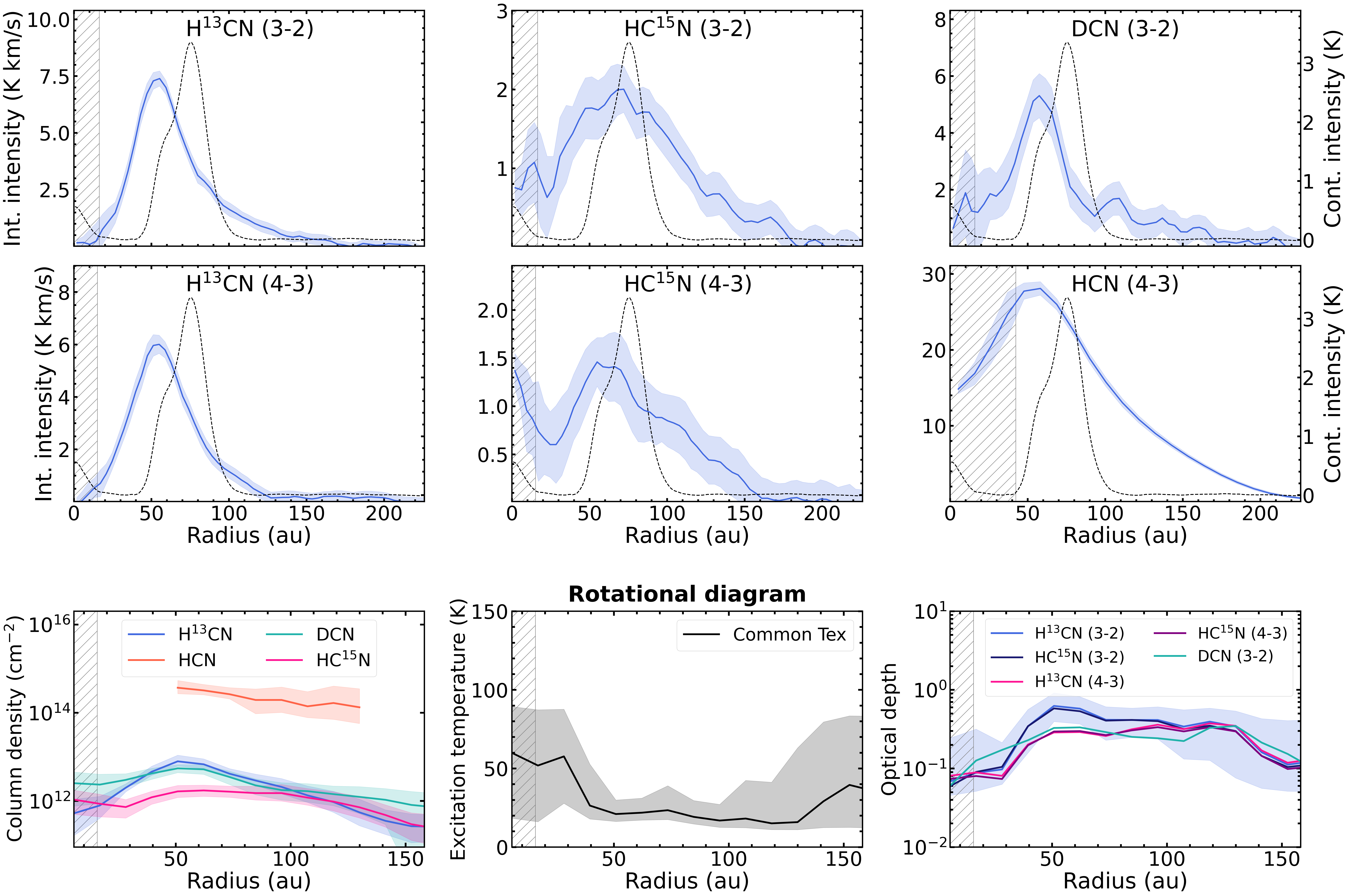}
    \caption{Same as Fig.~\ref{fig:profiles_rotdiagr}, but with cubes imaged to match the beam size of the HCN cube obtained from SB+LB observations. The radial profile of the integrated intensity of HCN is extracted from the cube obtained only from SB observations.}
    \label{fig:profiles_rotdiagr_b7res}
\end{figure*}

We could not smooth the HCN cube to match the spatial resolution of \ce{H^13CN} and \ce{HC^15N} Band 6 observations, as a worse spatial resolution would deteriorate the hyperfine fitting procedure (see also Sect.~\ref{subsec:HCN_column}). We therefore imaged the Band 6 \ce{H^13CN} and \ce{HC^15N} cubes again with Briggs weighting and a robust parameter of 0.67 to match the spatial resolution of Band 7 observations. We then applied again the rotational diagram analysis, as shown in Sect.~\ref{sec:column} and Fig.~\ref{fig:profiles_rotdiagr}: the new result of the column density profiles and excitation temperatures are shown in Fig.~\ref{fig:profiles_rotdiagr_b7res}. Similarly, we re-imaged the DCN cube with a robust parameter of 0.3 to match the beam size of Band 7 observations, and extract the column density profile. Figure~\ref{fig:12C13C} shows the corresponding fractionation profiles.

\FloatBarrier

\section{HCN emitting layer from the excitation temperature} \label{appendix:layer_HCN}

\begin{figure*}[ht!]
    \centering
    \includegraphics[scale=0.4]{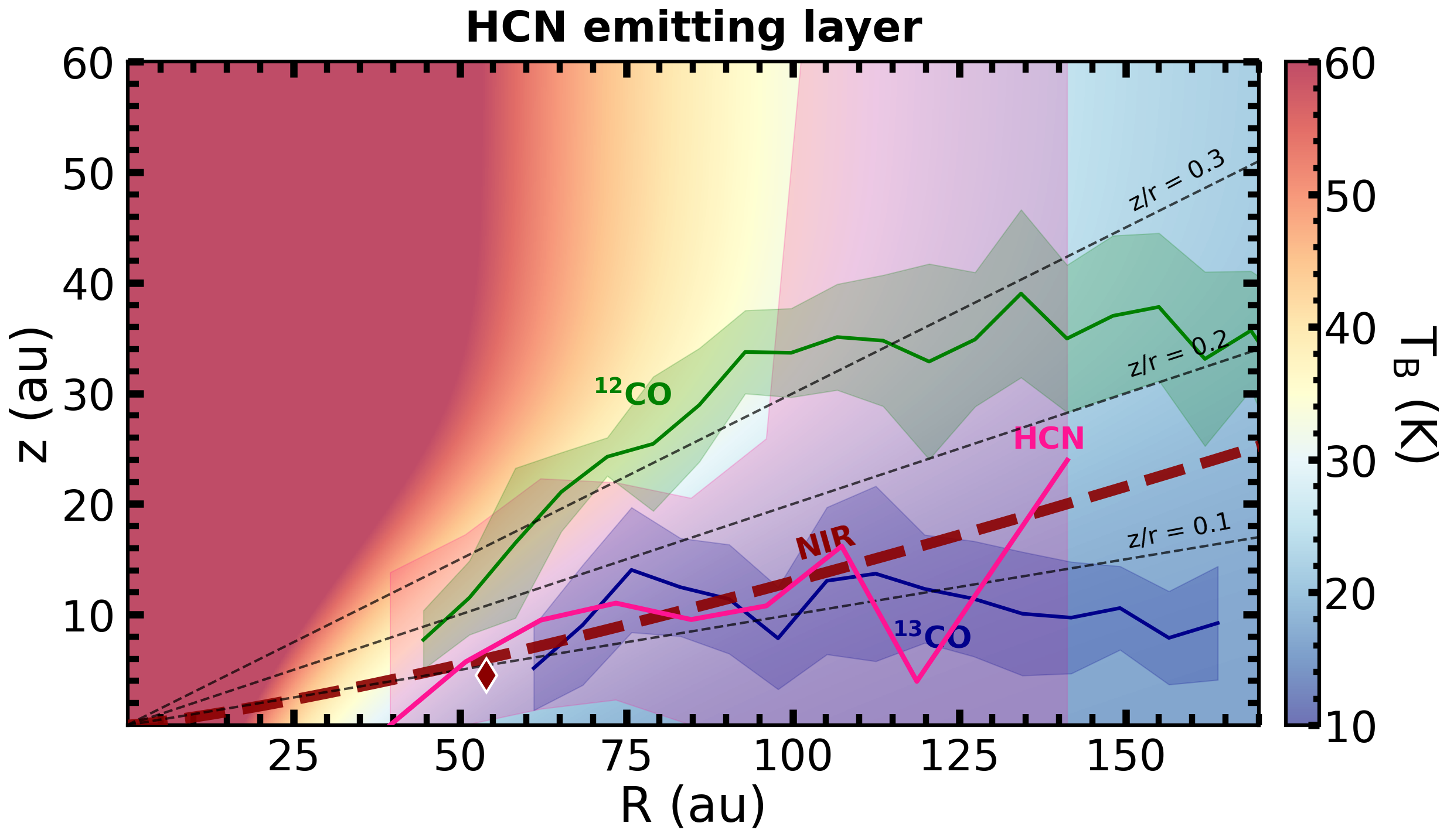}
    \caption{\ce{HCN} emitting layer height as a function of radius (solid pink line), with the related uncertainty (pink ribbons), compared to the 2D temperature structure in the background \citep{law2024mapping}.
    The two solid green and blue lines refer to the \ce{^12CO} and \ce{^13CO} emission surfaces, with related uncertainties, respectively \citep{law2024mapping}.  The dashed red line shows the NIR scale height as a function of radius, from \cite{keppler2018discovery}, while the red diamond is the height of the NIR emitting layer at the NIR ring $\sim0\farcs48$ \citep{law2024mapping}. Dashed black lines indicate fixed $z/r$ layers.}
    \label{fig:layer_hcn}
\end{figure*}

As we mentioned in Sect.~\ref{subsec:disc:N}, the distribution of the $\mu$m dust can influence the efficiency of isotope selective photodissociation of \ce{N2}, as it attenuates UV photons. It could be therefore important to constrain the emitting layer height of HCN with respect to the small dust.  We tried to retrieve the \ce{HCN} emitting layer, by comparing the radial profile of the excitation temperature in Fig.~\ref{fig:profiles_rotdiagr} with the 2D temperature structure of the PDS~70 disk, extracted by \cite{law2024mapping}, from CO isotopologues and \ce{HCO^+}. The rotational diagram assumes that the excitation temperature $T_\mathrm{ex}(r)$ is equal to the kinetic temperature $T(r,z)$ (LTE). We therefore inverted the $T(r,z)$ relation \citep{law2024mapping} to extract the emitting layer $z(r)$ \citep[see also][]{ilee2021molecules, rampinelli2024}, assuming a geometrically thin emitting layer, and that the HCN excitation temperature is the same as the one obtained for \ce{H^13CN} and \ce{HC^15N}. 
The result is presented in Fig.~\ref{fig:layer_hcn} (pink line), but the large uncertainties make it inconclusive.

\FloatBarrier

\section{HCN fractionation across the timeline of star and planet formation}
\label{appendix:other_stages}

A summary of the hydrogen, carbon, and nitrogen fractionation of HCN in various astronomical objects at different stages of star and planet formation is presented in Fig.~\ref{fig:frac_stages}.

\textit{\underline{DCN/HCN}}: The solar and ISM isotopic ratios are obtained from atomic hydrogen \citep{lodders2003sun, linsky2006}. Hot core DCN/HCN ratios were extracted from observations taken with the Arizona Radio Observatory Submillimeter Telescope (SMT) \citep[][HMC009.62, HMC010.47, HMC029.96, HMC045.47, HMC075.78, NGC7538B, Orion-KL, W3IRS5, W3H2 O]{gerner2015deuteration}. Column densities were extracted by correcting for the line optical depth, assuming the best-fit excitation temperature obtained from a 1D physico-chemical model with time-dependent D-chemistry \citep{gerner2015deuteration}. We highlight that these results are in tension with DCN/HCN ratios previously extracted for the same sources by \cite{hatchell1998}, from DCN and \ce{HC^15N} observations taken with the James Clerk Maxwell Telescope, which result in DCN/HCN values up to approximately an order of magnitude lower than the ones presented by \cite{gerner2015deuteration}. The five DCN/HCN values presented for class 0/I sources \citep[][Ser-emb 1,7,8,15,17]{bergner2020deuteration} were obtained from ALMA observations, evaluating HCN column densities from \ce{H^13CN}, assuming an ISM \ce{^12C/^13C}, $T_\mathrm{ex}=30$~K, and co-spatial emission from the different isotopologues. Disk integrated values of DCN/HCN were extracted for the protoplanetary disks of AS~209, IM~Lup, V4046~Sgr, LkCa~15, and HD~163296 from ALMA observations of \ce{H^13CN} and DCN \citep{huang2017alma}. The only DCN/HCN ratio measured for comets was obtained from James Clerk Maxwell Telescope observations of the Hale-Bopp comet \citep{meier1998deuteration}. The deuteration level was measured from the line ratios of DCN and \ce{H^13CN}, assuming a solar \ce{^12C/^13C}, but also directly from the main isotopologue HCN, by correcting for its optical depth through the hyperfine components, leading to consistent results.

\textit{\underline{HCN/\ce{H^13CN}}}: The ISM isotope ratio was obtained from atomic carbon \citep{wilson1999isotopes}, while the solar value was obtained from CO \citep{lyons2018carbon}. HCN/\ce{H^13CN} ratios were extracted for the prestellar core Bernard~1b \citep{daniel2013}, L1498 \citep{magalhaes2018nitrogen}, the six starless and prestellar cores CB23, TMC2, L1495, L1495AN, L1512, and L1517B \citep{jensen2024fractionation} and the class 0 YSO L483 \citep{agundez2019carbon} from IRAM observations. The HCN/\ce{H^13CN} value for the disk around TW~Hya was extracted from ALMA observations \citep{hily-blant2019multiple}. Cometary HCN/\ce{H^13CN} ratios were obtained for Hale-Bopp \citep{bockelee2008HCNcomets} from Harlan J. Smith Telescope observations, and for C/2012 S1 ISON \citep{cordiner2019carbon} from ALMA observations leveraging the hyperfine components of the observed HCN line.

\textit{\underline{HCN/\ce{HC^15N}}}: The solar and ISM isotopic ratios were obtained from atomic nitrogen \citep{marty2012earth, ritchey2015}. The prestellar/protostellar values are taken from \cite{hily-blant2020}. The first direct measurement of HCN/\ce{HC^15N} was extracted for the prestellar core L1498 \citep{magalhaes2018nitrogen} and subsequently for the six starless and prestellar cores CB23, TMC2, L1495, L1495AN, L1512, and L1517B \citep{jensen2024fractionation}, using a 1D radiative transfer code handling hyperfine overlap. The remaining HCN/\ce{HC^15N} values for prestellar cores L183, L1544 \citep{hily-blant2013nitrogen}, L1521E \citep{Ikeda2002nitrogen}, protocluster OMC-2 FIR4 \citep{evans2022fractionation}, and class 0/I YSOs FIR3 \citep{evans2022fractionation}, L1527 \citep{yoshida2019nitrogen}, IRAS 16293A, OMC-3, R CrA IRS7B \citep{wampfler2014nitrogen}, I-04365, I-04016, I-04166, I-04169, I-04181 \citep{legal2020nitrogen}, Ser-emb 1, 7, 8, 15, 17 \citep{bergner2020deuteration} were obtained from a so-called double-isotope indirect method, assuming a fixed \ce{^12C/^13C} ratio. The only direct measurement of the HCN/\ce{HC^15N} ratio in protoplanetary disks was obtained for the disk around TW~Hya \citep{hily-blant2019multiple}, while for the disks around AS~209, LkCa~15, and V4046~Sgr it was obtained from ALMA observations of \ce{H^13CN} and \ce{HC^15N} and assuming an ISM \ce{^12C/^13C} ratio \citep{guzman2017}. Cometary HCN/\ce{HC^15N} were presented for 17P/Holmes and Hale-Bopp \citep{bockelee2008HCNcomets} from IRAM, Harlan J. Smith, and Keck telescopes observations, and leveraging the hyperfine components of the observed HCN line.

\end{appendix}

\end{document}